\documentclass[a4paper,11pt]{article}
\pdfoutput=1 

\usepackage{jcappub} 
\usepackage[T1]{fontenc} 

\usepackage{graphicx,amssymb,amsmath,amsthm,amsfonts,epsfig,times,bm}
\usepackage{aas_macros}
\usepackage{cleveref}
\usepackage{tensor}
\usepackage{subfigure}

\def\be{\begin{equation}}
\def\ee{\end{equation}}
\def\bea{\begin{eqnarray}}
\def\eea{\end{eqnarray}}

\def\no{\nonumber}

\title{Thermodynamic stability of black holes surrounded by quintessence}

\author[1,2]{Meng-Sen Ma,}
\author[1,2]{Ren Zhao}
\author[3]{Ya-Qin Ma}
\affiliation[1]{Department of Physics, Shanxi Datong
University,  Datong 037009, China}
\affiliation[2]{Institute of Theoretical Physics, Shanxi Datong
University, Datong 037009, China}
\affiliation[3]{Medical College, Shanxi Datong University, Datong 037009, China}

\emailAdd{mengsenma@gmail.com; ms\_ma@sxdtdx.edu.cn}

\abstract{%
We study the thermodynamic stabilities of uncharged and charged black holes surrounded by quintessence (BHQ) by means of effective thermodynamic quantities.
When the state parameter of quintessence $\omega_q$ is appropriately chosen, the structures of BHQ are something like that of black holes in de Sitter space.
Constructing the effective first law of thermodynamics in two different ways, we can derive the effective thermodynamic quantities of BHQ.
Especially, these effective thermodynamic quantities  also satisfy Smarr-like formulae.
It is found that the uncharged BHQ is always thermodynamically unstable due to negative heat capacity, while for the charged BHQ there are phase transitions of the second order. We also show that there is a great deal of difference on the thermodynamic properties and critical behaviors of BHQ between the two ways we employed.}

\begin{document}

\maketitle
\flushbottom

\tableofcontents


\section{Introduction}

Since the Hawking-Page phase transition for Schwarzschild-AdS black hole was proposed, fruitful phase structures for different black holes in AdS space have been found. Specifically, the thermodynamic critical phenomena of charged AdS black hole have been studied extensively. It is shown in \cite{Chamblin-1999,Chamblin-1999a,Wu-2000,Banerjee-2012a}
 that the critical behaviors of Reissner-Nordstr\"{o}m-AdS black hole are analogous to that of Van de Waals liquid/gas system.
 Recently, it is found that the cosmological constant should be viewed as a thermodynamic variable --- pressure, and its conjugate quantity as the thermodynamic volume\cite{Wang-2006,Kastor-2009,Kastor-2010,Dolan-2011,Dolan-2011a}.
In the extended phase space,  $P-V$ criticality and other critical behaviors of several black holes are reexamined\cite{Kubiznak-2012,Gunasekaran-2012,Zhao-2013,Wei-2013,Chen-2013,Zou-2014,Ma-2014,Dehghani-2014,Zhang-2015}.
For a review, see\cite{Altamirano-2014}. Recently, this idea is also generalized to study the thermodynamic properties and critical behaviors of black holes in de Sitter space\cite{Zhao-2014a,Zhang-2014,Ma-2014b}.

It is widely believed that our Universe is accelerating expansion due to some unknown dark energy with negative pressure. The cosmological constant is a possible candidate. However, there are also other candidates, such as quintessence, phantom and quintom. In this paper, we only concern with the existence of quintessence, with which black hole spacetime should be modified.
Kiselev first studied black holes surrounded by quintessence (BHQ)\cite{Kiselev-2003}. Soon, quasinormal modes\cite{Chen-2005,Zhang-2006}, thermodynamic properties\cite{Chen-2008,Wei-2011,Azreg-Aienou-2013} and $P-V$ criticality\cite{Li-2014} of BHQ were extensively studied.

Similar to black holes in de Sitter space, BHQ can also has multiple horizons, at least a black hole event horizon and an outer horizon from quintessence (for short, we call it quintessence horizon). Generally, the temperatures on the two horizons are different, which makes the whole spacetime cannot be in thermodynamic equilibrium.
To overcome this problem, one can analyze the two horizons separately and independently. For instance, one can analyze one horizon and take another one as the boundary or separate the two horizons by a thermally opaque membrane or wall\cite{Wang-2002,Gomberoff-2003,Sekiwa-2006,Saida-2009}.  Besides,  one can also take a global view to construct the globally effective temperature and other effective thermodynamic quantities by analogy with the first law of thermodynamics\cite{Urano-2009,Zhao-2014a,Zhang-2014,Ma-2014b}.
Even though adopting the globally effective approach to deal with BHQ, there are still two different starting points.
One can first take the volume between the black hole horizon and the quintessence horizon as the thermodynamic volume of the system and then derive the effective pressure and other effective thermodynamic quantities from the the effective first law. In this case, the effective first law can be written in the form $dM=\tilde{T}_{eff}dS-P_{eff}dV+\cdots$, where the ``$\cdots$" represents the contributions from other matter fields. This idea is, in fact,  the generalization of that in\cite{Urano-2009,Zhao-2014a,Zhang-2014,Ma-2014b}. The another idea is to consider the quintessence parameter $\alpha$ as a thermodynamic variable and then derive other effective thermodynamic quantities. The effective first law in this case has the form $dM=T_{eff}dS+\Theta_{eff}d\alpha+\cdots$. Obviously, in the two cases, the effective thermodynamic quantities are different.

In fact, to study the thermodynamic stability of the black holes with multiple horizons in the \emph{global} sense non-equilibrium thermodynamics is needed. Until now, there are several works on the thermodynamic properties of black holes in de Sitter space. But, none of these deals with the \emph{global} thermodynamic stability. According to the effective thermodynamic quantities, we can define  heat capacities, free energy, etc. With these thermodynamic quantities, thermodynamic stability and phase transition of the thermodynamic system can be analyzed. As far as we know, the method of effective thermodynamic quantities is the unique one to analyze the thermodynamic properties of black holes with multiple horizons in a global view.

The paper is arranged as follows. In Section 2, we will introduce the black holes surrounded by quintessence with or without electric charge. In Section 3 we will study the the effective thermodynamic quantities and corresponding critical behaviors by considering the parameter $\alpha$ as the thermodynamic variable. In section 4, we do the similar things to that in section 3, but considering the geometry between the two horizons as a thermodynamic variable. We  make some concluding remarks in Section 5.

\section{Black holes surrounded by quintessence (BHQ)}

The metric of spherically symmetric charged black hole surrounded by quintessence is
\begin{equation}\label{metric}
ds^2=-f(r)dt^2+f(r)^{-1}dr^2 + r^2d\Omega^2,
\end{equation}
where
\be\label{generalmetric}
f(r)=1 - \frac{2 M}{r} + \frac{Q^2}{r^2} - \frac{\alpha}{r^{3\omega_q+1}} ,
\ee
and $M$ is the mass parameter, $Q$ is the electric charge, $\alpha$ is a normalization factor and $\omega_q$ is the state parameter of quintessence which is confined in the range $-1<\omega_q<-1/3$. Besides, the parameter $\alpha$ is related to the energy density, $\rho_q=-\frac{\alpha}{2}\frac{3\omega_q}{r^{3(1+\omega_q)}}$. Because $\rho_q$ is positive, $\alpha$ must take \emph{positive} values.

In the following, we will choose $\omega_q=-2/3$ for simplicity. In this case, Eq.(\ref{generalmetric}) becomes
\be\label{metric}
f(r)=1 - \frac{2 M}{r} + \frac{Q^2}{r^2} - \alpha r.
\ee

Firstly, we consider the uncharged BHQ. For such a spacetime,  $f(r)=0$  has two positive real roots:
\be
r_{+}=\frac{1-\sqrt{1-8 \alpha  M}}{2 \alpha }, \qquad r_{q}=\frac{1+\sqrt{1-8 \alpha  M}}{2 \alpha },
\ee
when $8 \alpha M \leq 1$.

Clearly, when $M=1/8 \alpha$ the two roots coincide. The smaller root $r_{+}$ corresponds to the black hole event horizon. And we call the larger one $r_{q}$, " quintessence horizon ", which is similar to the cosmological horizon in de Sitter space.

The parameters $M,~\alpha$ can be represented according to $r_{+}$ and $r_q$:
\be
M=\frac{x r_q}{2 (x+1)}, \qquad \alpha=\frac{1}{(x + 1)r_q},
\ee
where we have set $x=r_+/r_q$, thus $0\leq x \leq 1$.

The surface gravities of black hole horizon and the quintessence horizon are:
\be\label{ksds}
\kappa_{+}=\frac{1-2 \alpha  r_+}{2 r_+}>0, \quad \kappa_{q}=\frac{1-2 \alpha  r_q}{2 r_q}<0,
\ee
and
\be
 \kappa_{+}>|\kappa_{q}|.
\ee
Temperatures on both horizons are
\be
T_{+}=\frac{\kappa_{+}}{2\pi}, \qquad T_{q}=-\frac{\kappa_{q}}{2\pi}.
\ee
And entropies for the two horizons are respectively:
\be
S_{+}=\pi r_{+}^2, \qquad S_{q}=\pi r_{q}^2.
\ee

When considering the cosmological constant as a thermodynamic variable in de Sitter black holes, the first law of thermodynamics
 and associated Smarr formulae can be established for black hole horizon and cosmological horizon\cite{Dolan-2013a}.
 In analogy with black holes in de Sitter space, it is found that the first laws of thermodynamics can also be established on the two horizons if one takes $\alpha$ as a thermodynamic variable:
\be\label{1st}
dM=\frac{\kappa_{+}}{2\pi}dS_{+}+\Theta_{+}d\alpha, \quad
dM=\frac{\kappa_{q}}{2\pi}dS_{q}+\Theta_qd\alpha,
\ee
where $\Theta_{+}$ and $\Theta_{q}$ are the conjugate quantities to $\alpha$. There are also Smarr-like formulae for the two horizons:
\be
M=2T_{+}S_{+}-\Theta_{+}\alpha, \quad  M=-2T_{q}S_{q}-\Theta_{q}\alpha.
\ee

Now it is time to consider the charged BHQ. From Eq.(\ref{metric}), the horizons of the charged BHQ are determined by the cubic equation:
\be
\alpha r^3-r^2+2Mr-Q^2=0.
\ee
For this equation, there is a discriminant $\Delta$, which is
\be
\Delta=-32 \alpha  M^3+4 M^2+36 \alpha  M Q^2-27 \alpha ^2 Q^4-4 Q^2.
\ee

\begin{figure}[ht]
\begin{center}
\epsfig{file=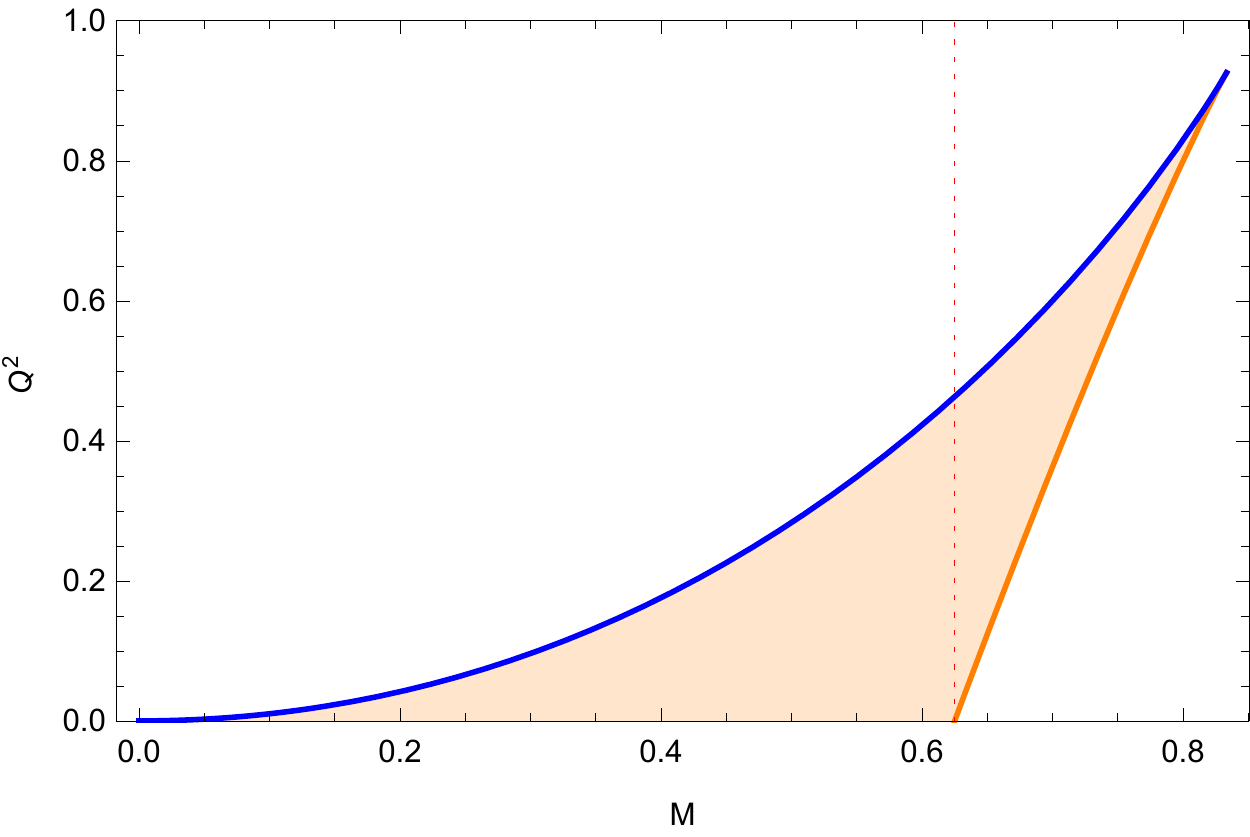,width=0.48\textwidth,angle=0,clip=true}
\caption{
Relations between $Q^2$ and $M$ for nonnegative discriminant $\Delta$. Here we set $\alpha=1/5$. The dotted (red) line lies at $M=1/8\alpha$. The borders intersect at $M=1/6\alpha$.
\label{figQM}}
\end{center}
\end{figure}

\begin{figure}[ht]
\begin{center}
\subfigure[]{\epsfig{file=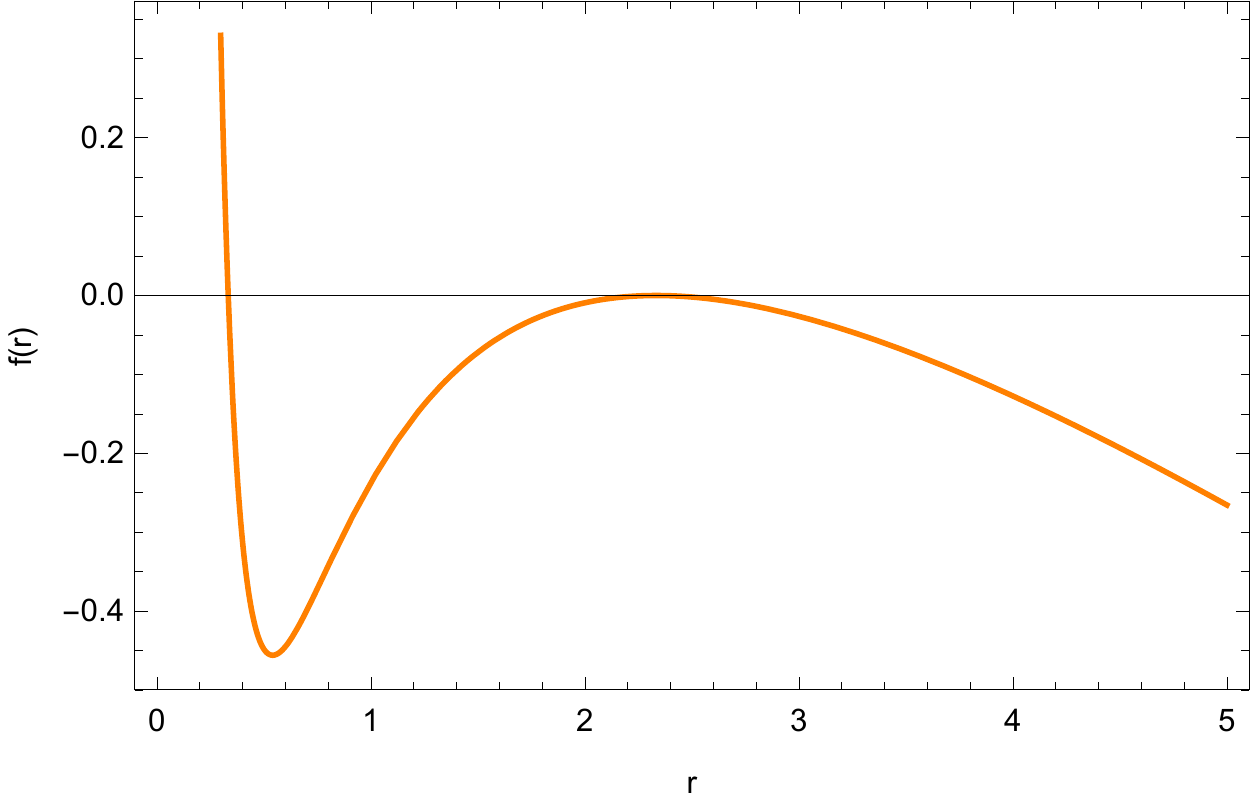,width=0.4\textwidth,angle=0,clip=true} \hspace{0.5cm}}
\subfigure[]{\epsfig{file=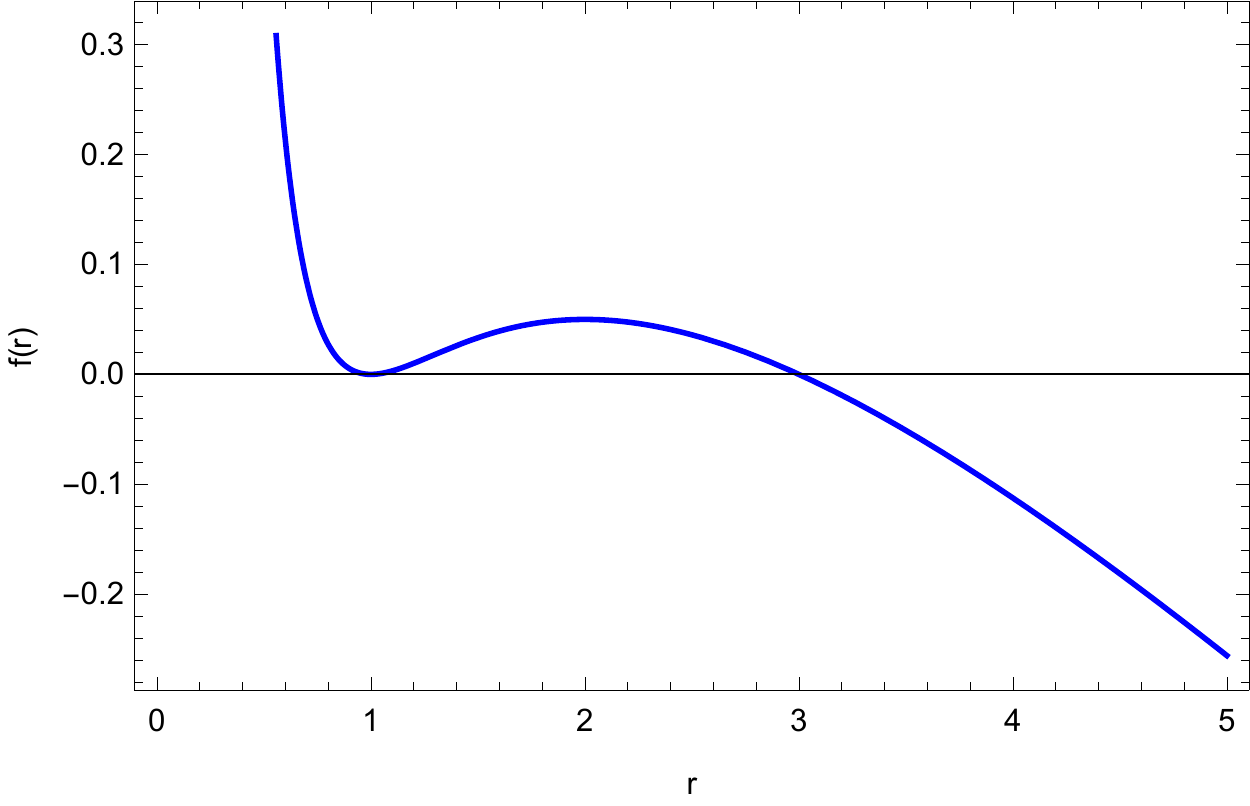,width=0.4\textwidth,angle=0,clip=true}}
\caption{
The two degenerate cases: (a) Nariai black hole; (b) cold black hole \label{figfr}}
\end{center}
\end{figure}

If $\Delta >0$, the equation has three distinct real roots, if $\Delta = 0$, the equation has one real root and one degenerate root, and if $\Delta<0$, complex roots arise, which are physically meaningless.
Below we only concern with the cases with $\Delta \geq 0$.
As is depicted in Figure \ref{figQM}, for fixed $\alpha$ only when the values of ($M, ~Q^2$) lie in the shadow region, the discriminant $\Delta$ can be positive and the borders of the shadow region correspond to
$\Delta=0$. The shadow region, in fact, can also be divided into two subregions according to $0<M<1/8\alpha$ and $1/8\alpha<M<1/6\alpha$. In the two subregions, the behaviors of solutions have a great difference. Figure \ref{figfr} shows the two degenerate case. When the black hole event horizon coincides with the quintessence horizon, it is called Nariai black hole. When the black hole inner horizon and the event horizon coincide, it is called cold black hole. When the three horizons coincide, it is ultrcold black hole. This has been discussed in detail in \cite{Fernando-2013}.

\begin{figure}[ht]
\begin{center}
\epsfig{file=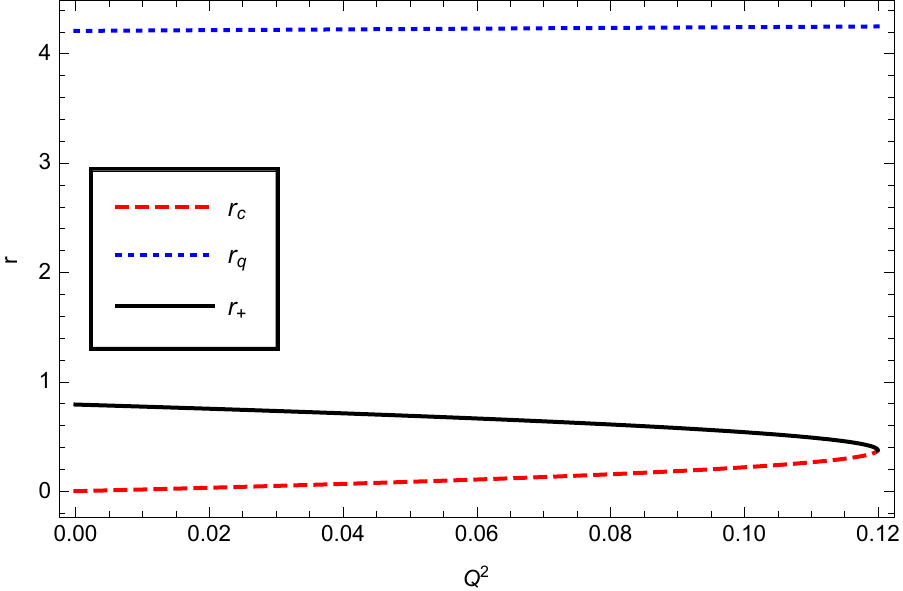,width=0.4\textwidth,angle=0,clip=true} \hspace{0.5cm}
\epsfig{file=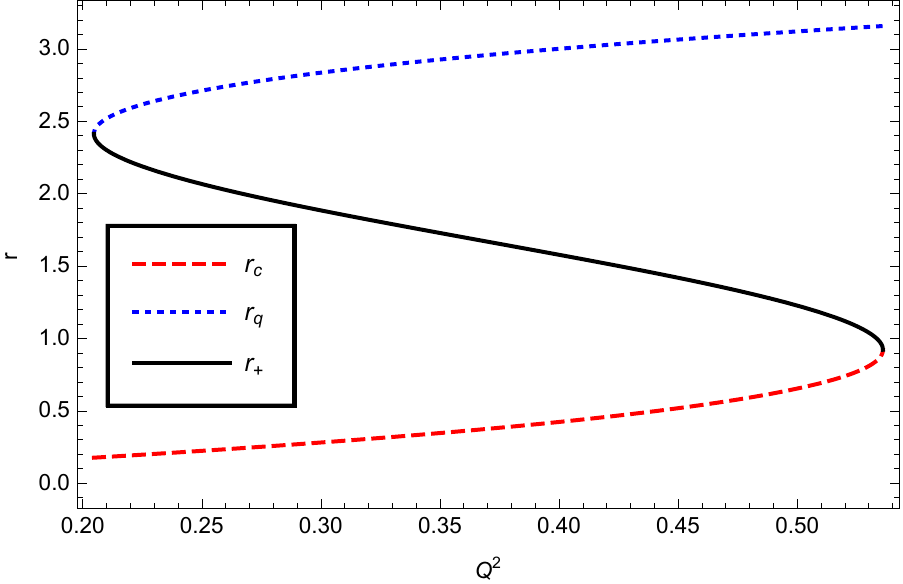,width=0.4\textwidth,angle=0,clip=true}
\caption{The three roots $r_{c}, ~r_{+},~r_{q}$ as functions of $Q^2$ in two different cases. The left subfigure corresponds to the case with $M<1/8\alpha$ and we take $M=1/3$, $\alpha=1/5$. The right subfigure corresponds to the case with $1/8\alpha<M<1/6\alpha$ and we take $M=2/3$, $\alpha=1/5$.
\label{figrQ}}
\end{center}
\end{figure}

The explicit expressions of the three roots when $\Delta>0$ are given by
\bea
r_{c}&=&\frac{1}{3 \alpha }-\frac{2 p}{3} \cos \left(\frac{\theta }{3}\right), \\
r_{+}&=&\frac{1}{3 \alpha }+\frac{2 p}{3} \cos \left(\frac{\theta }{3}+\frac{\pi }{3}\right), \\
r_{q}&=&\frac{1}{3 \alpha }+\frac{2 p}{3} \cos \left(\frac{\theta }{3}-\frac{\pi }{3}\right),
\eea
where we have taken
\be
p=\frac{\sqrt{1-6 \alpha  M}}{\alpha }, \quad h= \frac{18 M \alpha - 27 Q^2 \alpha^2 - 2}{\alpha^3}, \quad \cos\theta=\frac{h}{2p^3}.
\ee
There is a relation between $(p,~h,~\Delta)$: $4p^6\alpha^2=\alpha^2h^2+27\Delta$, which means multiple root arises when $h=\pm 2p^3$.
It is found that the three roots are all positive  only if $6 \alpha  M <1$.
Among the three roots, the smallest one, $r_c$, corresponds to the black hole Cauchy/inner
horizon, the intermediate one, $r_{+}$, corresponds to the black hole event horizon, and the largest one, $r_{q}$, corresponds to the quintessence horizon.
In Figure \ref{figrQ}, we show the relations between the three roots and $Q^2$ for fixed $M$ and $\alpha$. The sizes of $r_{c}, ~r_{+},~r_{q}$ are also clearly shown.

For the charged case, according to $r_{+},~r_{q}$ and $Q$, we can express $M$ and $\alpha$ as
\be
M=\frac{x^2 r_q^2+Q^2 x^2+Q^2 x+Q^2}{2 x (x+1) r_q}, \qquad \alpha=\frac{x r_q^2-Q^2}{x (x+1) r_q^3}.
\ee
When the black hole horizon and the quintessence horizon are viewed as independent each other, there are also respective first laws of thermodynamics:
\be\label{1st}
dM=\frac{\kappa_{+}}{2\pi}dS_{+}+\Phi_{+}dQ+\Theta_{+}d\alpha, \quad
dM=\frac{\kappa_{q}}{2\pi}dS_{q}+\Phi_{q}dQ+\Theta_qd\alpha,
\ee
where $\Phi_{+}$ and $\Phi_{q}$ are electric potentials corresponding to the two horizons. There are also Smarr-like formulae for the two horizons:
\be
M=2T_{+}S_{+}+\Phi_{+}Q-\Theta_{+}\alpha, \quad  M=-2T_{q}S_{q}+\Phi_qQ-\Theta_{q}\alpha,
\ee
where the minus sign before $T_{q}S_q$ is also due to the negative surface gravity for the quintessence horizon.

It can be found that although the BHQ is very different from Reissner-Nordstr\"{o}m-dS (RNdS) black hole, the behaviors of their horizons are very similar. Below we will continue to analyze thermodynamic properties of BHQ by means of the global and effective method. When taking the global view, we have two choices. We can certainly take the parameter $\alpha$ as a thermodynamic variable as we did above. Besides, we can also take the volume $V$ between the black hole horizon and the quintessence horizon as a thermodynamic volume. This idea has been extensively studied in de Sitter black holes\cite{Zhao-2014a,Zhang-2014,Ma-2014b,Urano-2009}.

Before proceeding, we should stress two key points. First, the method we employed to cope with BHQ is an effective method. As we mentioned above, the thermodynamic system of BHQ is not in thermodynamic equilibrium states in general, due to the different temperatures on the two horizons. Globally, we can just define effective thermodynamic quantities by analogy. Second, when considering the BHQ as a whole, what is the total entropy? We can take Schwarzschild-dS black hole as an example. As we all know, when the black hole horizon and the cosmological horizon exist separately, the entropies both satisfy the Bekenstein-Hawking area law. However, when the two horizons exist at the same time, the total entropy of the whole system may be not the sum of the entropies of the two horizons due to the mutual effects between the two horizons. However, there are also some reasons to support the area law of total entropy for de Sitter black holes\cite{Kastor-1993,Cai-1998,Wang-2002,Urano-2009,Bhattacharya-2016}. Until now, there is still no consensus on the entropy of de Sitter black holes. For BHQ, it has the similar problem. In this paper, for simplicity we take the second viewpoint on the total entropy to deal with BHQ.

\section{$\alpha$ as thermodynamic variable}

\subsection{Uncharged BHQ}

First we analyze the uncharged black hole surrounded by quintessence. One can introduce the effective thermodynamic quantities and construct effective thermodynamic law for the uncharged BHQ:
\be
dM=\frac{\kappa_{eff}}{2\pi}dS+\Theta_{eff}d\alpha,
\ee
where we have taken $S=S_{+}+S_{q}$ as the total entropy of the BHQ.

The effective temperature $T_{eff}$ and the quantity $\Theta_{eff}$ are respectively\footnote{Here $\kappa_{eff}$ is negative. Physically acceptable temperature should be positive, so we take the absolute value. }:
\be\label{SdSQT}
T_{eff}=-\frac{\kappa_{eff}}{2\pi}=\frac{1}{4 \pi  r_q(1+x)}=\frac{\alpha }{4 \pi } , \qquad
\Theta_{eff}=-\frac{1}{2} \left(x^2+x+1\right) r_q^2.
\ee
One can easily verify that these effective thermodynamic quantities also satisfy a Smarr-like formula:
\be\label{Smarr1}
M=-2T_{eff}S-\Theta_{eff}\alpha.
\ee

When $x\rightarrow 0$, the black hole horizon vanishes. The uncharged BHQ  becomes the free quintessence spacetime. In this case, one can easily find that
$T_{eff}=\frac{1}{4 \pi  r_q}$ and $S=S_{q}$.
 These are exactly thermodynamic quantities of free quintessence spacetime.

 When $x\rightarrow 1$, the black hole horizon and the quintessence horizon \emph{apparently} coincide, which is usually called Nariai limit\cite{Nariai-1999,Ginsparg-1983}. In this case, $r_{+}=r_{q}=1/2\alpha$, thus $T_{+}=T_{q}=0$. However, the effective temperature turns into
$T_{eff}=\frac{1}{8 \pi  r_q}$.
In fact, the two horizons are not really coincident. Not only that, the nonzero temperature in the Nariai limit
is consistent with the result of Bousso and Hawking, although the coefficient is a little different\cite{Bousso-1996,Fernando-2013a}.
From Eq.(\ref{SdSQT}), one can easily see that $T_{eff}$  monotonically decreases as the $x$ increases for fixed $r_q$. In the Nariai limit, it reaches the minimum.

To understand the thermodynamic stability of the uncharged BHQ, we can calculate the heat capacity by means of the effective thermodynamic quantities. We can define
\be
C_\Theta=\left.\frac{\partial M}{\partial T_{eff}}\right|_{\Theta_eff}=-2 \pi  \left(2 x^2+3 x+2\right) r_q^2 < 0.
\ee
While, $C_\alpha$ cannot be defined in the similar way, because constant $\alpha$ means constant $T_{eff}$.
According to the viewpoint of Davies\cite{Davies-1977}, divergences of heat capacity means the second-order phase transition happens there.  Thus, we conclude that the global uncharged BHQ is thermodynamically unstable and no phase transition arises.

\subsection{Charged BHQ}

For the charged black hole surrounded by quintessence, taking the electric charge $Q$ as well as $\alpha$ to be thermodynamic variables, we can derive the effective thermodynamic quantities and corresponding first law of black hole thermodynamics:
\be\label{eff1st}
dM = \frac{\kappa_{eff}}{2\pi}dS + {\Phi _{eff}}dQ + \Theta_{eff}d\alpha.
\ee
The effective surface gravity $\kappa_{eff}$, the effective electric potential $\Phi_{eff}$ and the conjugate quantity $\Theta_{eff}$ to $\alpha$ are respectively:
\bea
\frac{\kappa_{eff}}{2\pi}&=&\left.\frac{\partial{M}}{\partial{S}}\right|_{Q,\alpha}=-\frac{\left(Q^2 (x+2)-x r_q^2\right) \left(Q^2 (2 x+1)-x^2 r_q^2\right)}{4 \pi  x (x+1) r_q^3 \left(x^2 r_q^2-Q^2 \left(x^2+3 x+1\right)\right)},\\
\Phi_{eff}&=&\left.\frac{\partial{M}}{\partial{Q}}\right|_{S,\alpha}=-\frac{Q^3 (x+1)}{r_q \left(x^2 r_q^2-Q^2 \left(x^2+3 x+1\right)\right)},\\
\Theta_{eff}&=&\left.\frac{\partial{M}}{\partial{\alpha}}\right|_{S,Q}=\frac{r_q^2 \left(x^2 \left(x^2+x+1\right) r_q^2-Q^2 \left(x^4+3 x^3+3 x^2+3 x+1\right)\right)}{2 \left(Q^2 \left(x^2+3 x+1\right)-x^2 r_q^2\right)}.
\eea
When $Q=0$, charged BHQ should return back to the uncharged BHQ. One can easily check that the above quantities indeed degenerate to that in  Eq.(\ref{SdSQT}), only if we define the effective temperature of the RNdS black hole to be:
\be
T_{eff}=\frac{|\kappa_{eff}|}{2\pi}.
\ee
Also, these effective thermodynamic quantities satisfy the Smarr-like formula:
\be\label{Smarr2}
M=-2T_{eff}S+\Phi_{eff}Q-\Theta_{eff}\alpha.
\ee
It should be noted that if $\alpha$ is not viewed as a thermodynamic variable, the relation above cannot be established.

For the charged BHQ, it is meaningless to discuss the $x \rightarrow 0$, because it will not return to any known black hole.
In the $x \rightarrow 1$ limit, the charged Nariai black hole is obtained\cite{Fernando-2013}. In this case, $r_{+}=r_{q}=\rho$ and  $M, ~\alpha $  have simple forms:
\be
M=\frac{\rho ^2+3 Q^2}{4 \rho }, \quad \alpha=\frac{\rho ^2-Q^2}{2\rho ^3}.
\ee
The inner horizon lies at
\be
r_{-}=b=\frac{1}{\alpha }-2 \rho < \rho.
\ee
The effective temperature $T_{eff}$ in the charged Nariai case, is
\be
T_{eff}=\frac{(b-\rho )^2}{4 \pi  \rho  (\rho -2 b) (b+2 \rho )}\neq 0.
\ee
In the ultracold case, namely the three horizons coincide, one can easily find that $T_{eff}=0$.

Next, we continue to discuss the thermodynamic properties of charged BHQ. Without loss of generality, we will take the choice $r_q=1$ below. It is found that $T_{eff}$ is always negative when $Q^2>1/3$.
A negative temperature does not make any sense in black hole thermodynamics. Thus, below we will study the thermodynamic properties of charged BHQ with $0\leq Q^2 \leq 1/3$.

\begin{figure}[ht]
\begin{center}
\epsfig{file=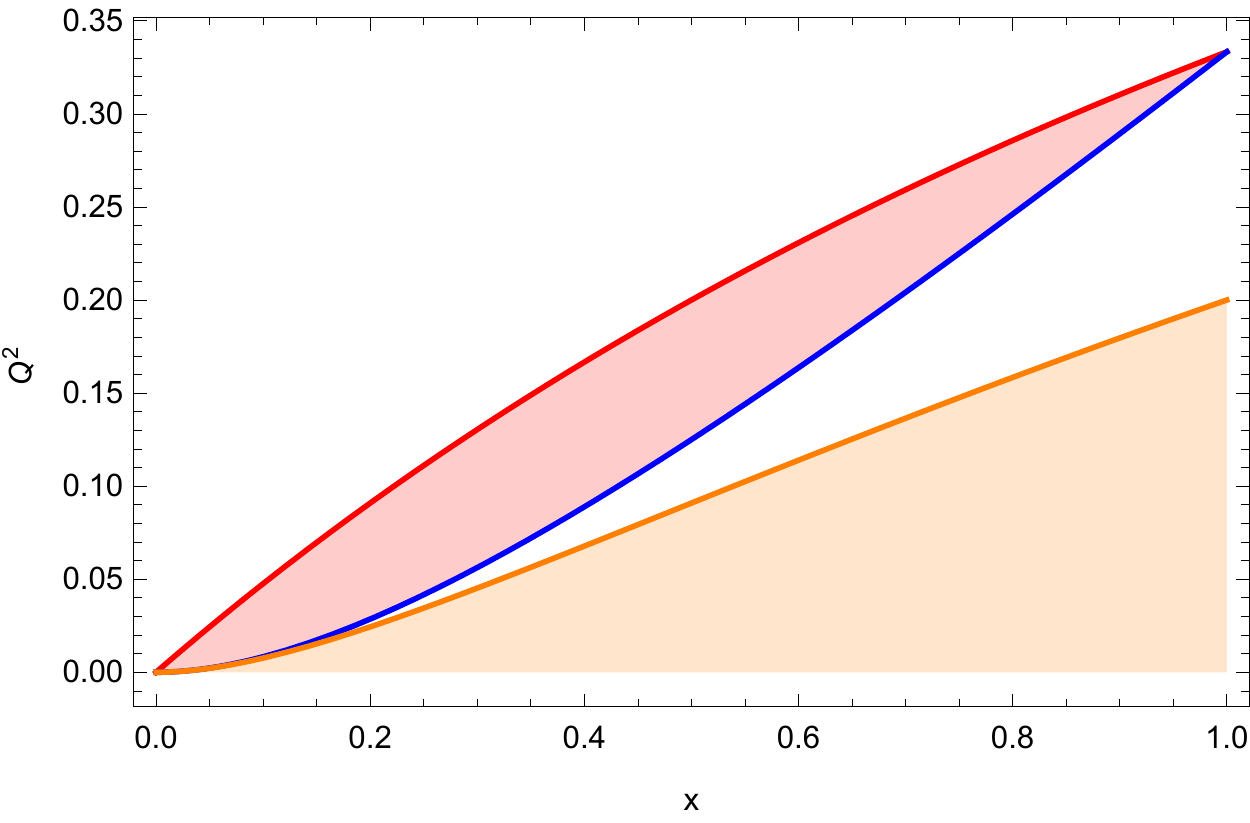,width=0.48\textwidth,angle=0,clip=true}
\caption{
The relations between $x$ and $Q^2$. Only in the shadow regions $T_{eff}$ is positive.
\label{figQx1}}
\end{center}
\end{figure}

\begin{figure}[ht]
\begin{center}
\subfigure[$Q=0.3$]{\epsfig{file=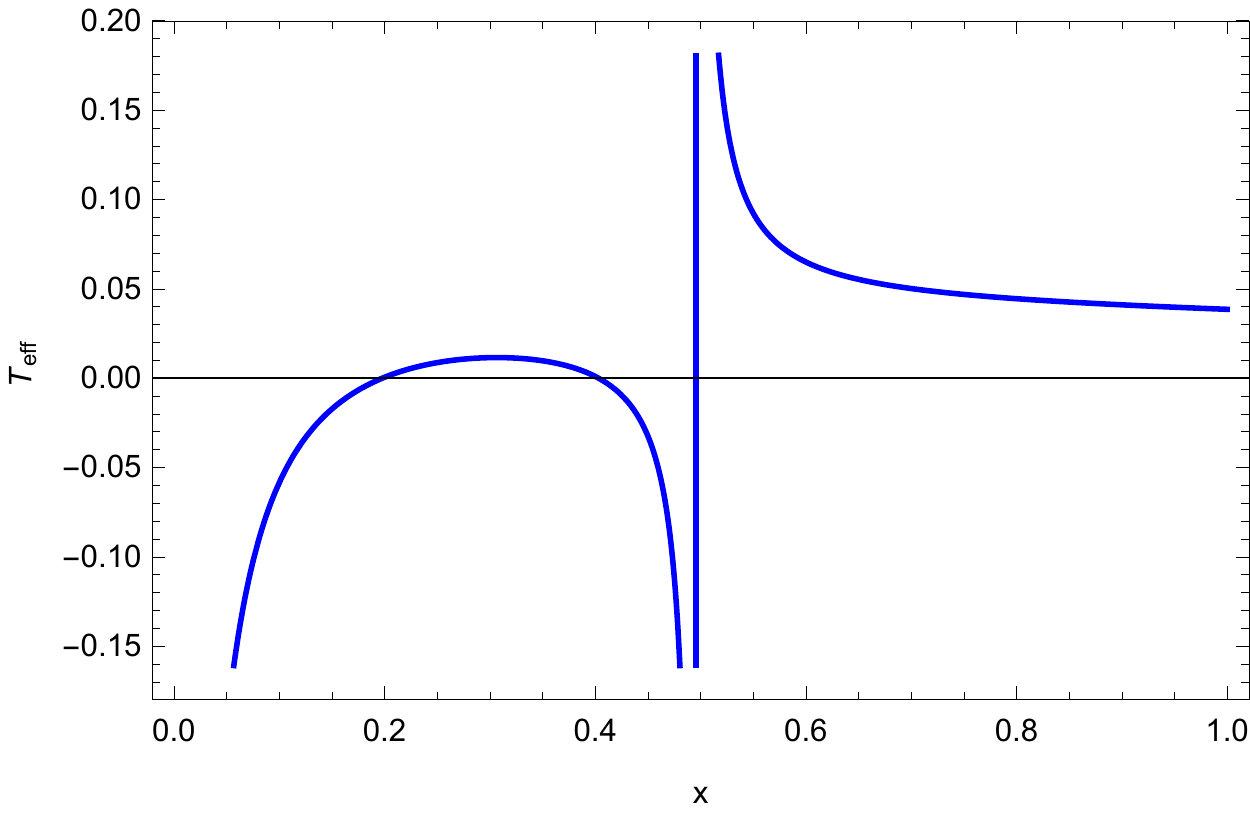,width=0.4\textwidth,angle=0,clip=true} \hspace{0.5cm}}
\subfigure[$Q=0.5$]{\epsfig{file=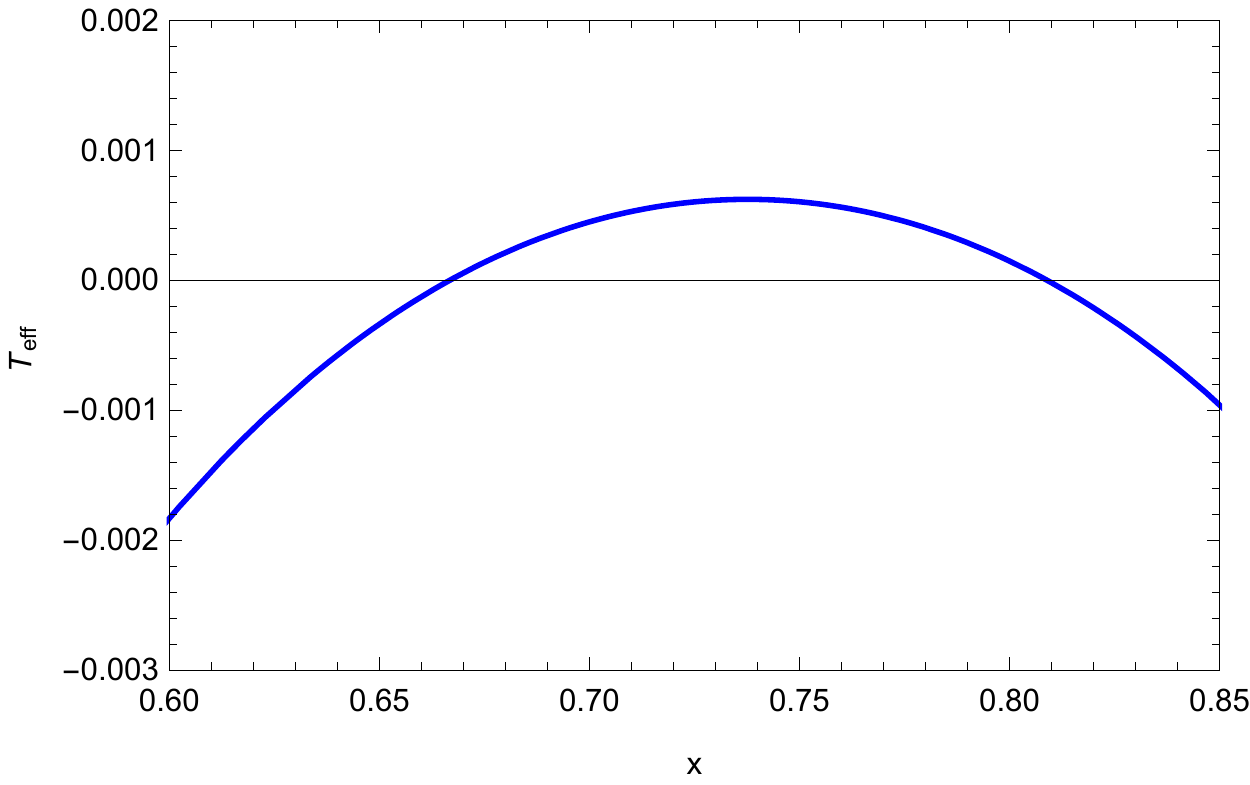,width=0.4\textwidth,angle=0,clip=true}}
\caption{The effective temperature $T_{eff}$ as functions of $x$. In (a), the two zero-temperature points lie at $x=0.198$ and $x=0.403$. The divergent point lies at $x=0.496$. In (b), the two zero-temperature points are $x=0.667$ and $x=0.809$.
\label{figQTx1}}
\end{center}
\end{figure}

To require $T_{eff}\geq 0$, there are some restrictions on $Q$ and $x$, which are shown in Fig.\ref{figQx1}.
For fixed $x$, there are always two regions in which the effective temperature $T_{eff}$ is positive. While for fixed $Q$, there will be one or two regions in which $T_{eff}$ is positive,  depending on the values of $Q$. As is shown in Fig.\ref{figQx1}, there is one region when $Q^2>1/5$ and  two regions when $Q^2 \leq 1/5$. Therefore, $T_{eff}$ has different behaviors for $Q^2 \in [0,1/5]$ and $Q^2 \in (1/5,1/3]$.
In Fig.\ref{figQTx1}, we plot the $T_{eff}$ for $Q=0.3$ and $Q=0.5$, respectively. In the first case, the effective temperature can be positive in two disconnected regions separated by a negative-temperature region. The charged BHQ cannot transit through a negative-temperature region. Thus, the charged BHQ can only stay in one of the two regions and will be always in that region.

\begin{figure}[ht]
\begin{center}
\subfigure[$Q=0.3$]{\epsfig{file=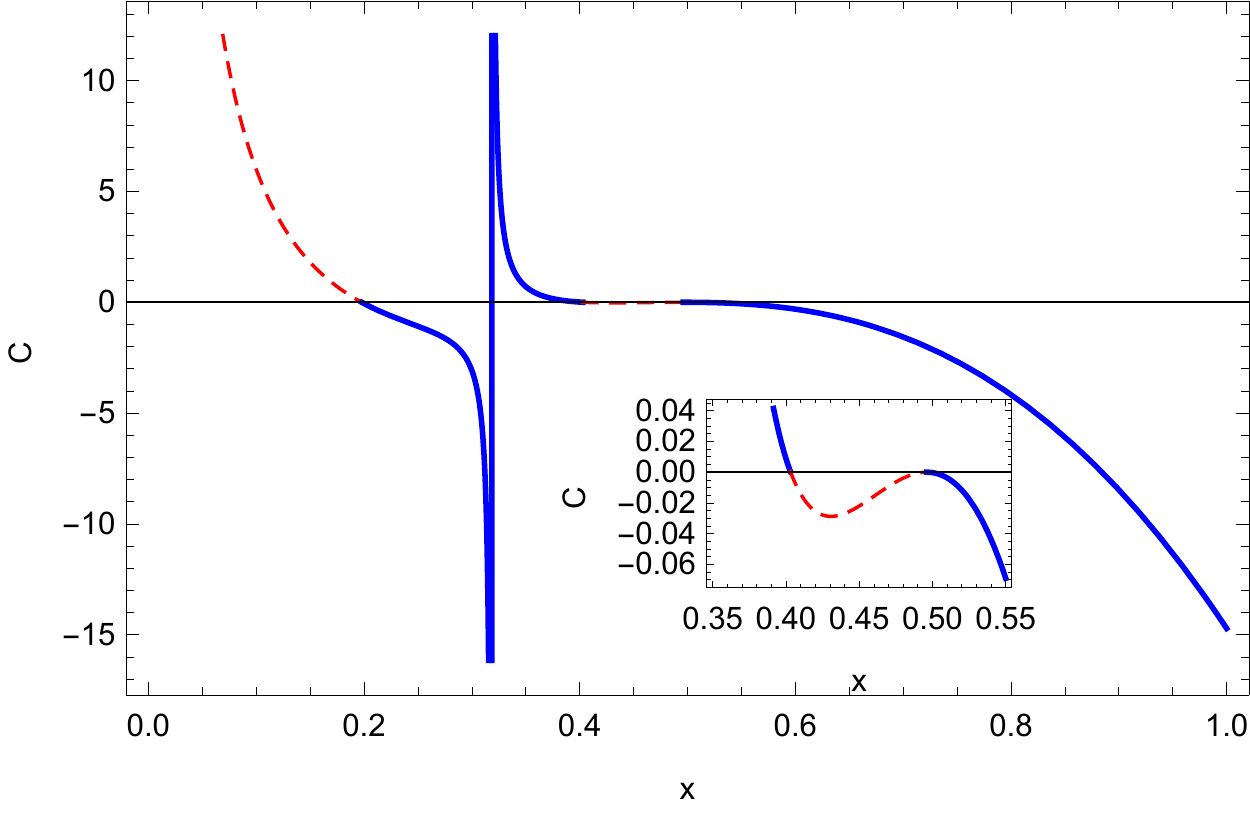,width=0.4\textwidth,angle=0,clip=true}} \hspace{0.5cm}
\subfigure[$Q=0.5$]{\epsfig{file=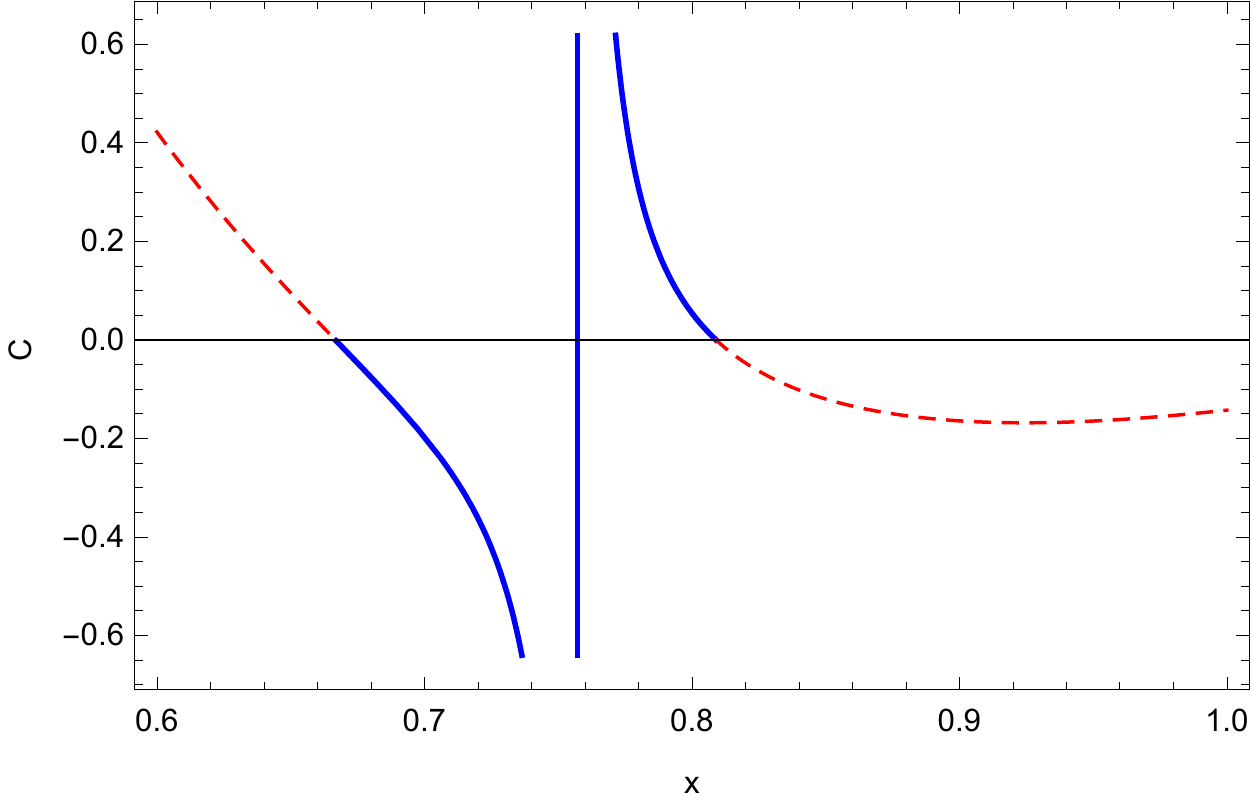,width=0.4\textwidth,angle=0,clip=true}}
\caption{The heat capacity $C$ as functions of $x$. The left subfigure corresponds to the case with $Q=0.3$. The right subfigure corresponds to the case with $Q=0.5$. The solid (blue) curves correspond to positive-temperature regions, and the dashed (red) curves correspond to the negative-temperature region.
\label{figcx1}}
\end{center}
\end{figure}

The heat capacity can be defined as
\be
C=\left.\frac{\partial{M}}{\partial{T_{eff}}}\right|_{Q,\alpha}=-\frac{2 \pi  \left(Q^2 (x+2)-x\right) \left(Q^2 (2 x+1)-x^2\right) \left(x^2-Q^2 \left(x^2+3 x+1\right)\right)^2}{Q^4 x (x+1)^2 \left(3 Q^4-Q^2 \left(x^2+8 x+1\right)+3 x^2\right)},
\ee
where we also have set $r_q=1$.
The behaviors of the heat capacity $C$ for $Q=0.3$ and $Q=0.5$ are depicted in Fig.\ref{figcx1}. In the two cases, there are both one divergent point, which corresponds to the point where $T_{eff}$ has local maximum values.
In the former case, the heat capacity has three zero points which correspond to the points of zero temperature and the point of divergent temperature, respectively. In the latter case, there are two zero points for the heat capacity, which are the points of zero temperature.

As is shown in Fig.\ref{figcx1}, for $Q=0.3$ case, the largest black hole with positive temperature has negative heat capacity, which means it is thermodynamically unstable.
Except for this part, heat capacities in the two cases have similar behaviors. The charged BHQ will transit from the smaller black hole with negative heat capacity to the larger one with positive heat capacity and a second-order phase transition happens at the divergent points of heat capacity.

To discuss the global stability of the charged BHQ,
we need to calculate the free energy, which is defined as
\bea
F&=& M-T_{eff}S \no \\
&=&\frac{Q^2 x \left(x^4+4 x^3-2 x^2+4 x+1\right)-Q^4 \left(4 x^4+13 x^3+14 x^2+13 x+4\right)+(x-1)^2 x^3}{4 x (x+1) \left(x^2-Q^2 \left(x^2+3 x+1\right)\right)}.
\eea

\begin{figure}[ht]
\begin{center}
\epsfig{file=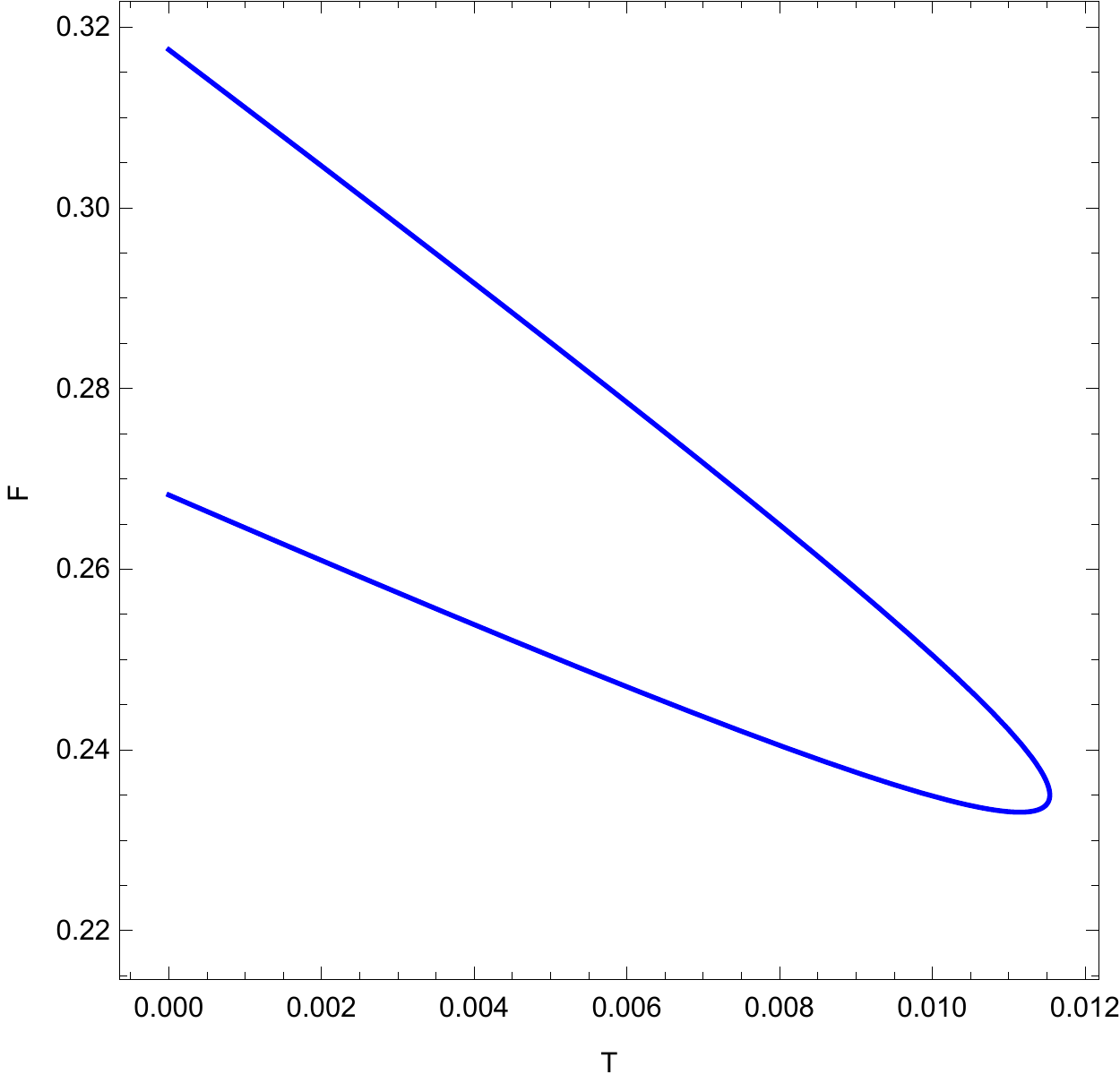,width=0.4\textwidth,angle=0,clip=true}
\caption{
The free energy as functions of $T_{eff}$ with $Q=0.3$.
\label{figFT}}
\end{center}
\end{figure}

Because the free energy has similar behavior in the two cases, we only plot the case with $Q=0.3$.  As is shown in Fig.\ref{figFT}, the lower branch corresponds to black holes with positive heat capacity, which are indeed thermodynamically stable.

\section{$V$ as thermodynamic variable}

In this section, we will take the volume between the black hole horizon and the quintessence horizon as a thermodynamic variable --- the thermodynamic volume, which has the form
\be
V=V_{q}-V_{b}=\frac{4\pi}{3} \left(1-x^3\right) r_q^3.
\ee
Besides, we also take the total entropy of the BHQ as $S=S_{+}+S_{q}$.

\subsection{Uncharged BHQ}

For the uncharged BHQ, the effective first law of thermodynamics is
\be
dM=\tilde{T}_{eff}dS-P_{eff}dV.
\ee

\be\label{VUCBHQ}
\tilde{T}_{eff}=\frac{x^4+1}{4 \pi  x (x+1)^3 r_q}, \quad P_{eff}=\frac{1-x^3}{8 \pi  x (x+1)^3 r_q^2}.
\ee
Clearly, these quantities are both always positive. Specifically, the effective temperature $\tilde{T}_{eff}$ is different from $T_{eff}$ in the former section. A Smarr-like formula is also established:
\be
M=2T_{eff}S-3P_{eff}V.
\ee
It is different from Eq.(\ref{Smarr1}) because we take different thermodynamic variable.

When $x \rightarrow 0$, the effective temperature and effective pressure diverge, thus are physically meaningless. In the $ x \rightarrow 1$ limit,
the effective temperature is $ \tilde{T}_{eff}=\frac{1}{16 \pi  r_q}$, which is different from the result in the former section. In this limit, the volume $V$ and the effective pressure $P_{eff}$ are zero.
However, this does not mean the two horizon coincide. In the Nariai case, the original coordinate is no longer appropriate to analyze the BHQ. In a new coordinate, it can be shown that the two horizons, in fact, are not degenerate\cite{Bousso-1996,Fernando-2013a}.

\begin{figure}[ht]
\begin{center}
\epsfig{file=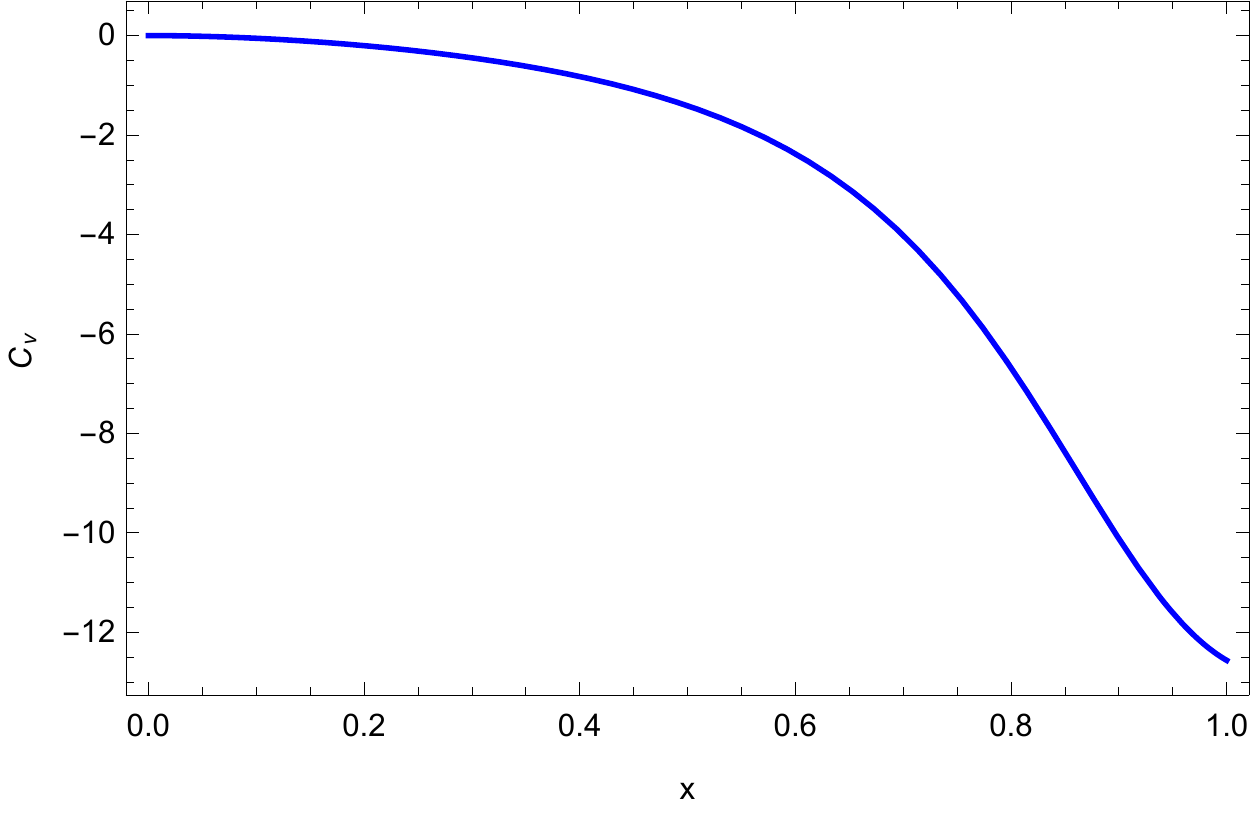,width=0.48\textwidth,angle=0,clip=true}
\caption{
The heat capacity $C_v$ as function of $x$.
\label{figcv}}
\end{center}
\end{figure}

The heat capacity at constant volume can be calculated
\be
C_{v}=\left.\frac{\partial M}{\partial \tilde{T}_{eff}}\right|_{V}=-\frac{2 \pi  x^2 (x+1)^2 \left(x^4+1\right) r_q^2}{x^8+4 x^7-6 x^4+4 x+1}.
\ee

As is shown in Fig. \ref{figcv}, it is always negative. For a thermodynamic system, to be stable equilibrium, it requires $C_P\geq C_V\geq 0$\cite{Callen-2006}. Therefore, we conclude that the uncharged BHQ is thermodynamically unstable. This result is consistent with that in the former section.

\subsection{Charged BHQ}

For the charged black hole surrounded by quintessence, we start from the effective first law:
\be
dM=\tilde{T}_{eff}dS+\tilde{\Phi}_{eff}dQ-P_{eff}dV,
\ee
where the effective temperature, the effective pressure and the effective electric potential are, respectively
\bea
\tilde{T}_{eff}&=&\frac{\left(x^6+x^2\right) r_q^2-Q^2 \left(x^6+2 x^5+2 x+1\right)}{4 \pi  x^3 (x+1)^3 r_q^3}, \\
P_{eff}&=&-\frac{x^2 \left(x^3-1\right) r_q^2-Q^2 \left(x^5+2 x^4-2 x-1\right)}{8 \pi  x^3 (x+1)^3 r_q^4},\\
\tilde{\Phi}_{eff}&=&\frac{Q \left(x^2+x+1\right)}{x (x+1) r_q}.
\eea
It can be checked that when $Q=0$, these quantities can return back to those in Eq.(\ref{VUCBHQ}). They satisfy the Smarr-like formula:
\be
M=2T_{eff}S-3P_{eff}V+\Phi_{eff}Q.
\ee

\begin{figure}[ht]
\begin{center}
\epsfig{file=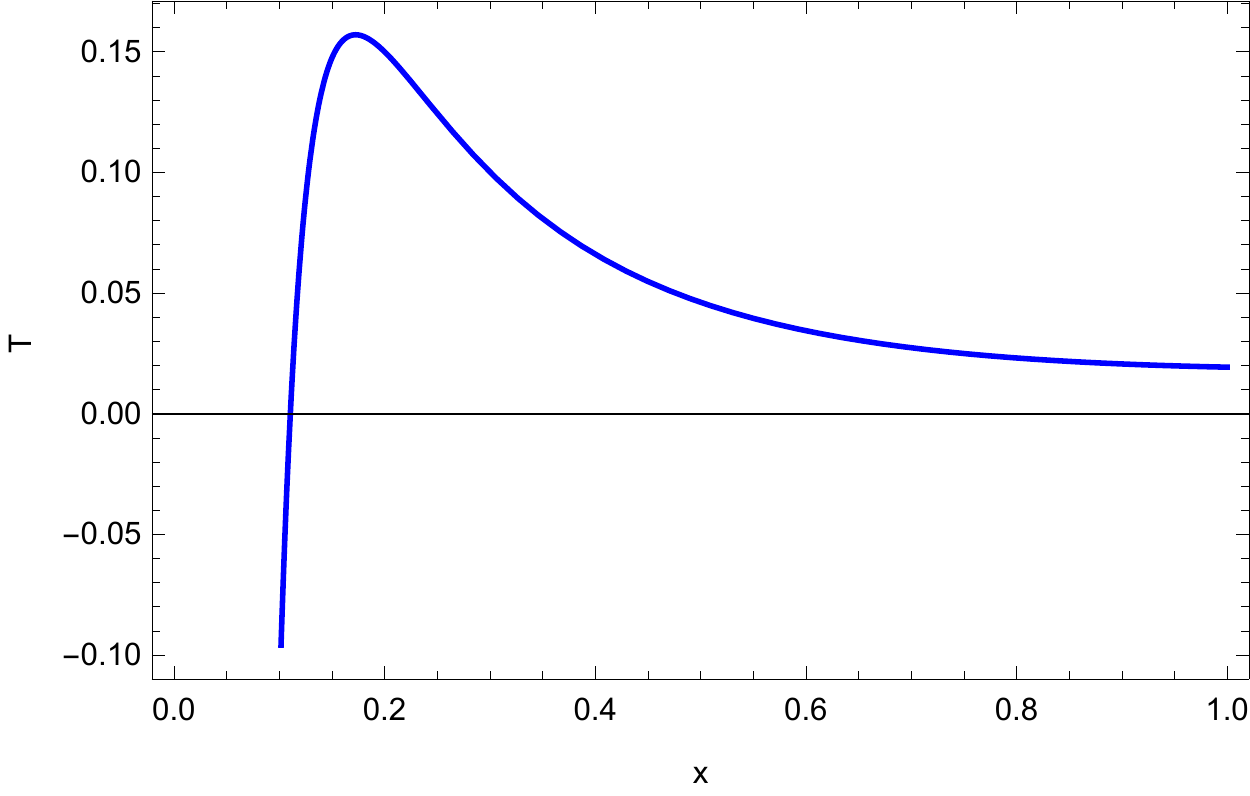,width=0.5\textwidth,angle=0,clip=true}
\caption{
The effective temperature $\tilde{T}_{eff}$ as function of $x$. Here we set $Q=0.1$. \label{figTTx}}
\end{center}
\end{figure}

We also set $r_q=1$ below. In this case, we find that when $Q^2\leq 1/3$, the behavior of $\tilde{T}_{eff}$ is something like that of RN black hole. We have depict $\tilde{T}_{eff}-x$ in Fig.\ref{figTTx}.

Now we can calculate the heat capacity of charged BHQ. We first consider heat capacity at constant volume, which is
\be\label{CVCBH}
C_v=\left.\frac{\partial{M}}{\partial{\tilde{T}_{eff}}}\right|_{Q,V}=\tilde{T}_{eff}\left.\frac{\partial{S}}{\partial{\tilde{T}_{eff}}}\right|_{Q,V}.
\ee
The complete expression is lengthy. We put it in the Appendix.

\begin{figure}[ht]
\begin{center}
\epsfig{file=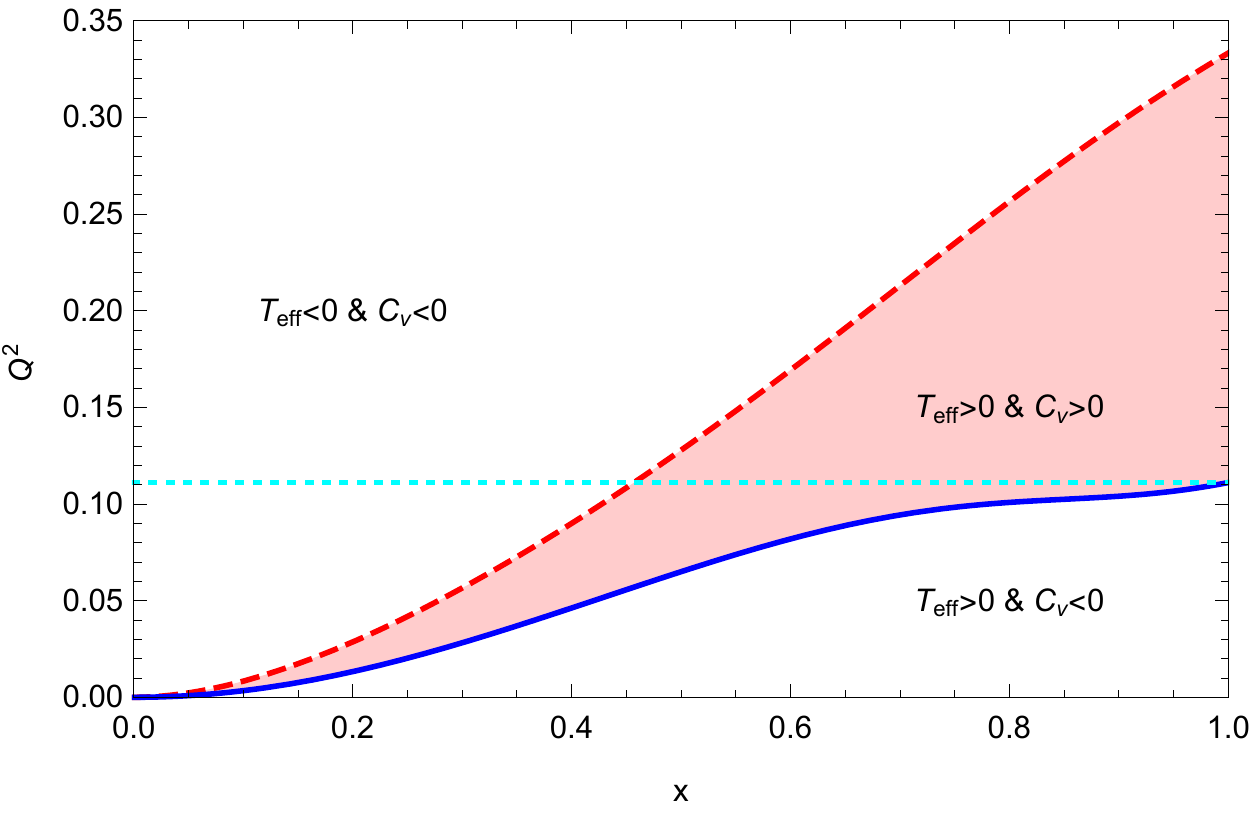,width=0.5\textwidth,angle=0,clip=true}
\caption{The signs of $\tilde{T}_{eff}$ and $C_v$ determined by $Q^2$ and $x$. We set $r_q=1$. The dashed (red) curve corresponds to zero temperature and zero heat capacity. The solid (blue) curve corresponds to the divergent point of heat capacity. The dashed curve ends at $Q^2=1/3$ and the solid curve ends at $Q^2=1/9$.
 \label{figcvQ2x}}
\end{center}
\end{figure}

\begin{figure}[ht]
\begin{center}
\epsfig{file=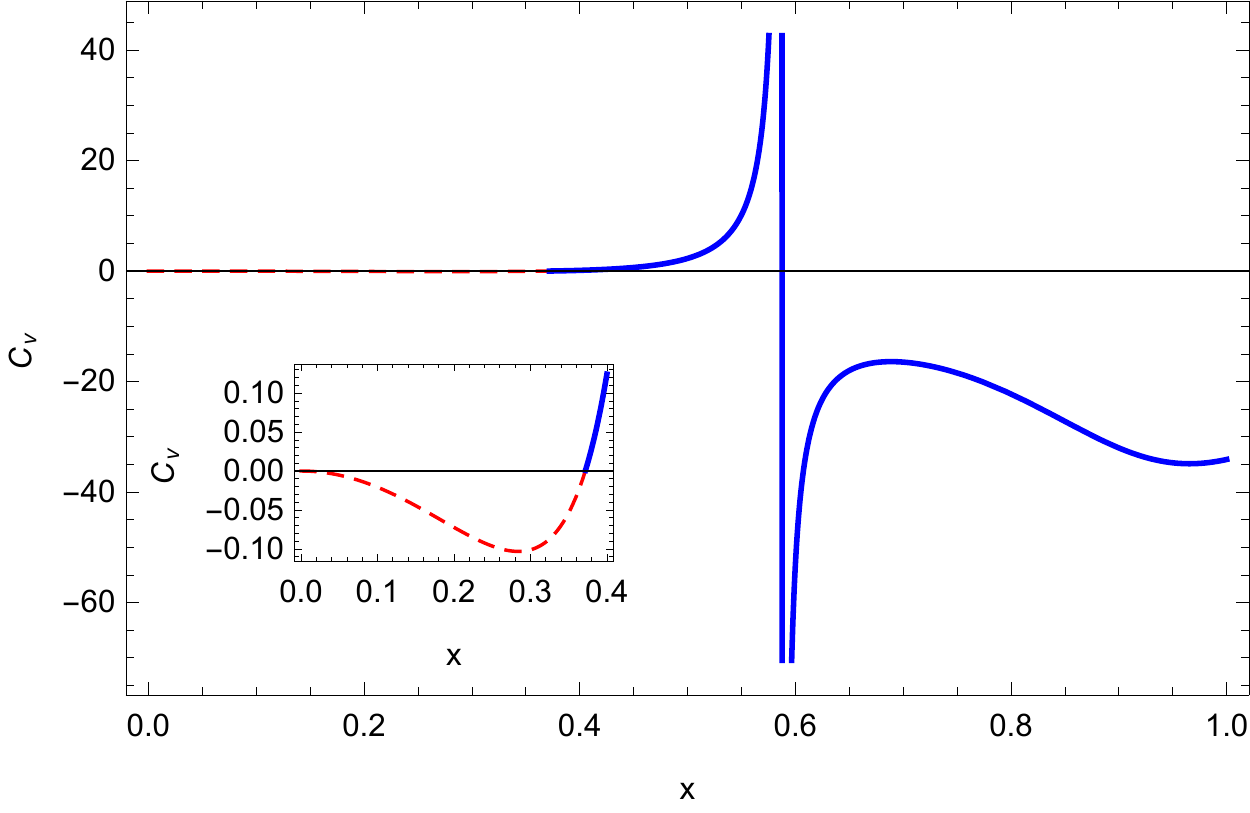,width=0.4\textwidth,angle=0,clip=true} \hspace{0.5cm}
\epsfig{file=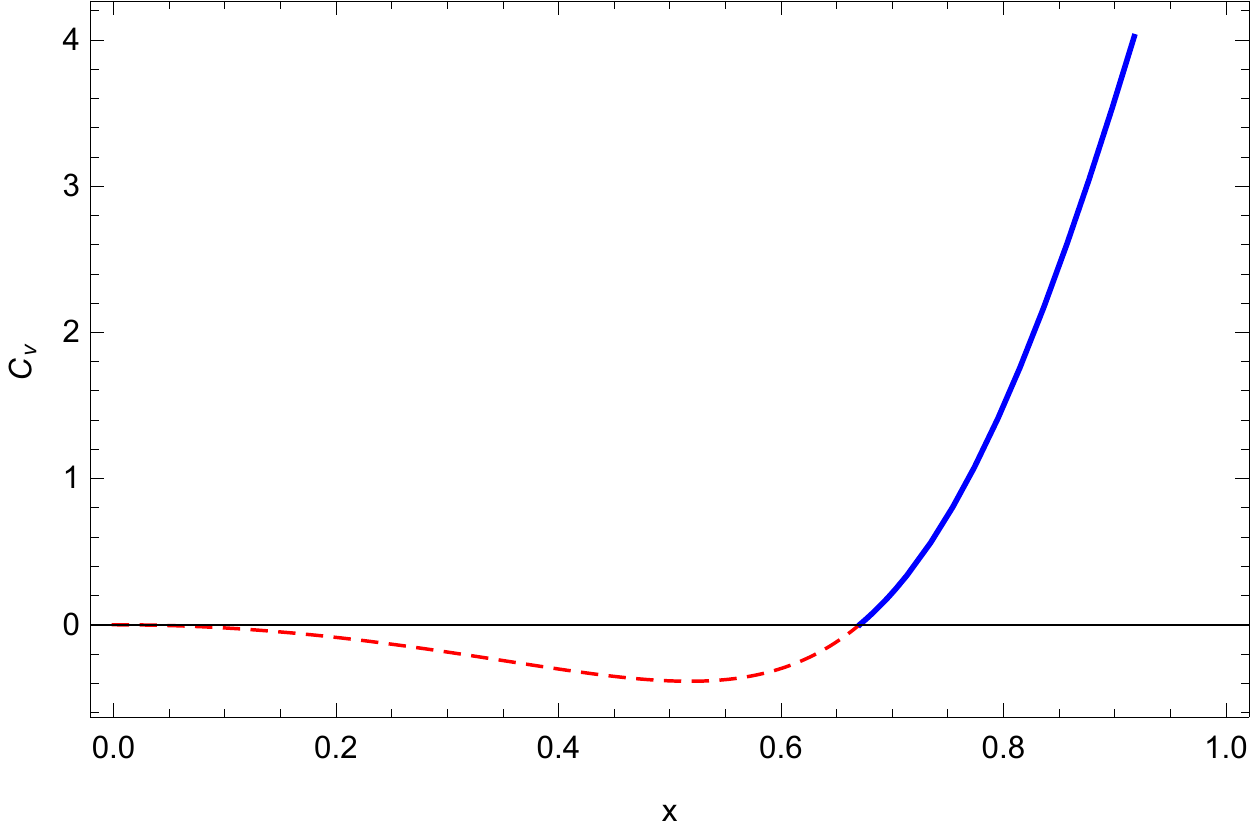,width=0.4\textwidth,angle=0,clip=true}
\caption{The heat capacity $C_v$ as functions of $x$. The left subfigure corresponds to the case with $Q=0.08$. The right subfigure corresponds to the case with $Q^2=0.2$. The solid (blue) curves correspond to positive-temperature regions, and the dashed (red) curves correspond to the negative-temperature region.
\label{figcvx}}
\end{center}
\end{figure}

When $Q=0$, it indeed degenerates to the result for the uncharged case. However, due to the existence of electric charge, $C_v$, not like the uncharged case,  can be positive.
From Eq.(\ref{CVCBH}), one can easily find that the zero points of $C_v$ and $\tilde{T}_{eff}$ are the same. As is shown in Fig.\ref{figcvQ2x}, the dashed curve which represents the points where $C_v$ and $\tilde{T}_{eff}$ are both zero and the solid curve which corresponds the divergent points of $\tilde{T}_{eff}$, divide the $Q^2-x$ figure into three regions. Only in the shadow region, the effective temperature and the heat capacity $C_v$ can be both positive and thus only in this region the charged BHQ is physically meaningful and thermodynamically stable. It should be noted that the shadow region can also be divided into two subregions by the horizontal line $Q^2=1/9$. Above the line, no divergence exists for $C_v$. Below the line, $C_v$ has a divergent point for fixed $Q$. In Fig.\ref{figcvx}, we show $C_v$ in the two cases.

\begin{figure}[ht]
\begin{center}
 \subfigure[$Q^2=0.06$]{\epsfig{file=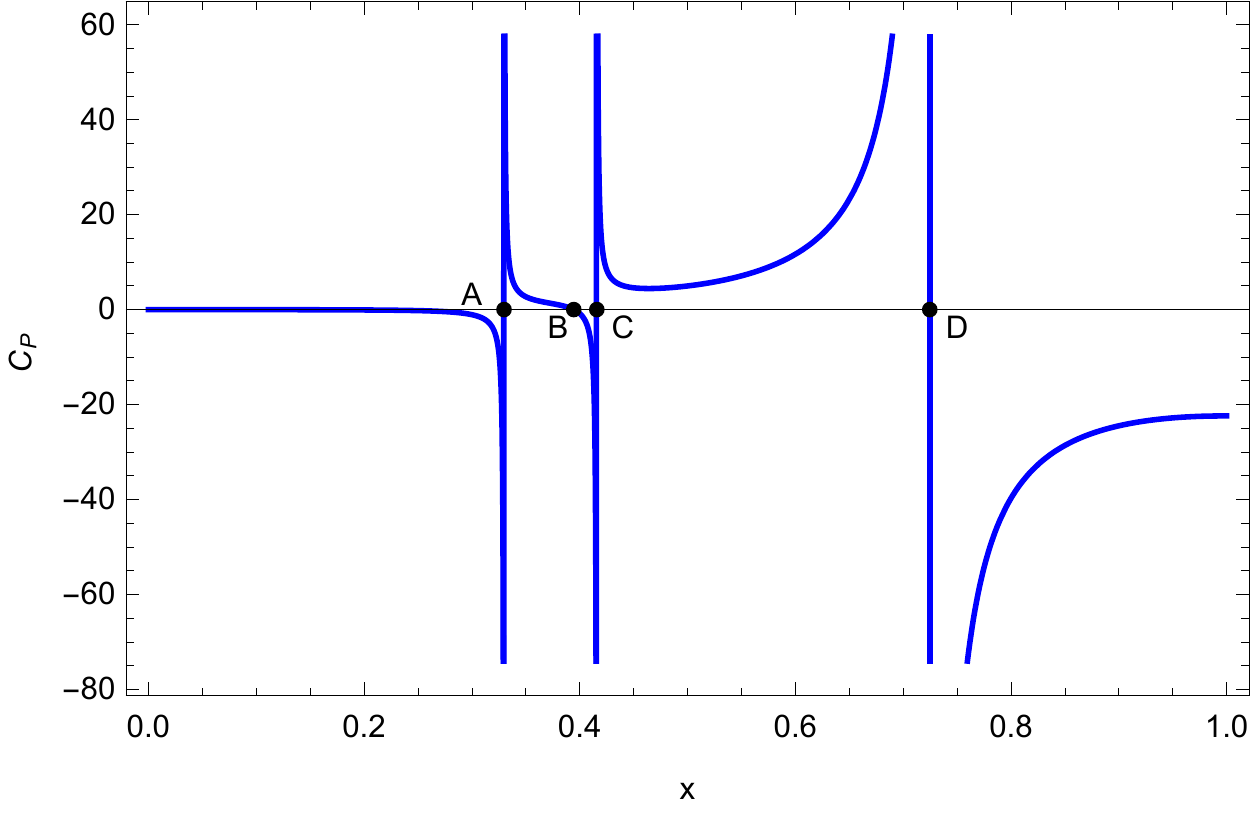,width=0.3\textwidth,angle=0,clip=true}}
 \subfigure[$Q^2=0.08$]{\epsfig{file=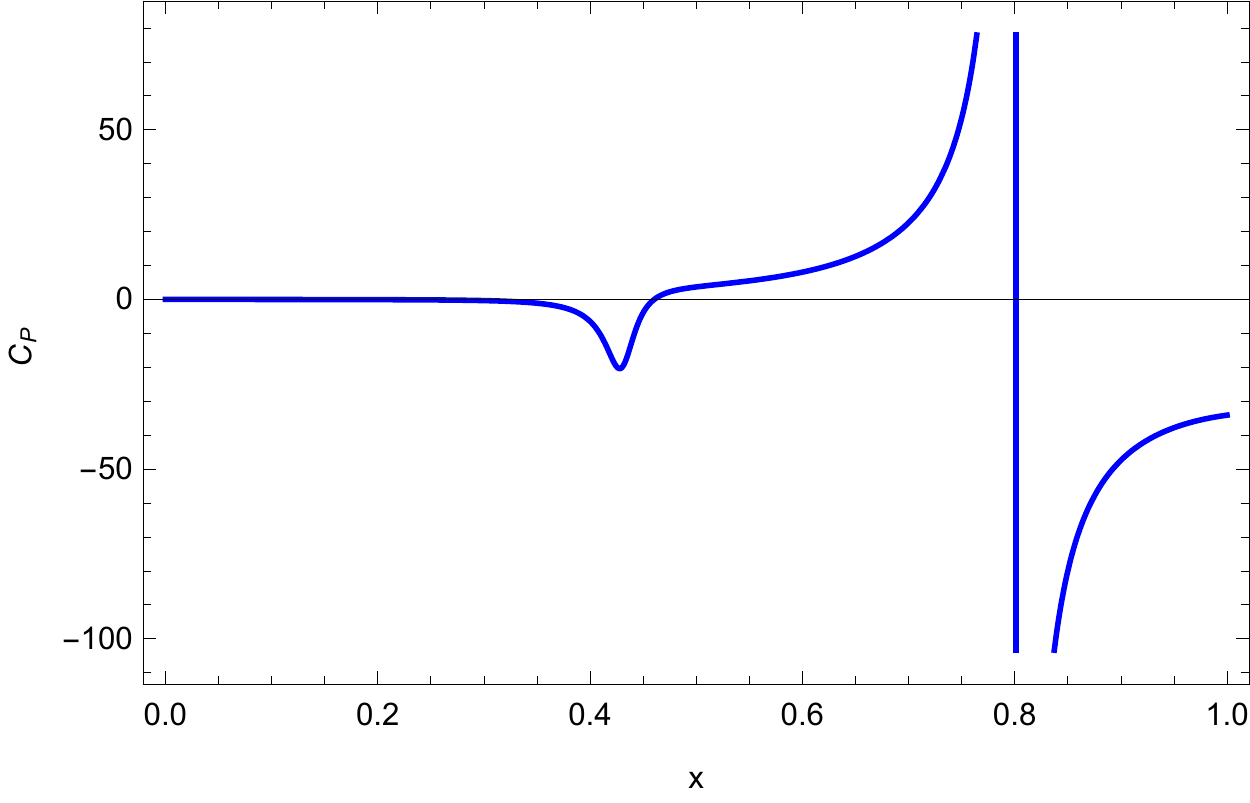,width=0.3\textwidth,angle=0,clip=true}}
 \subfigure[$Q^2=0.15$]{\epsfig{file=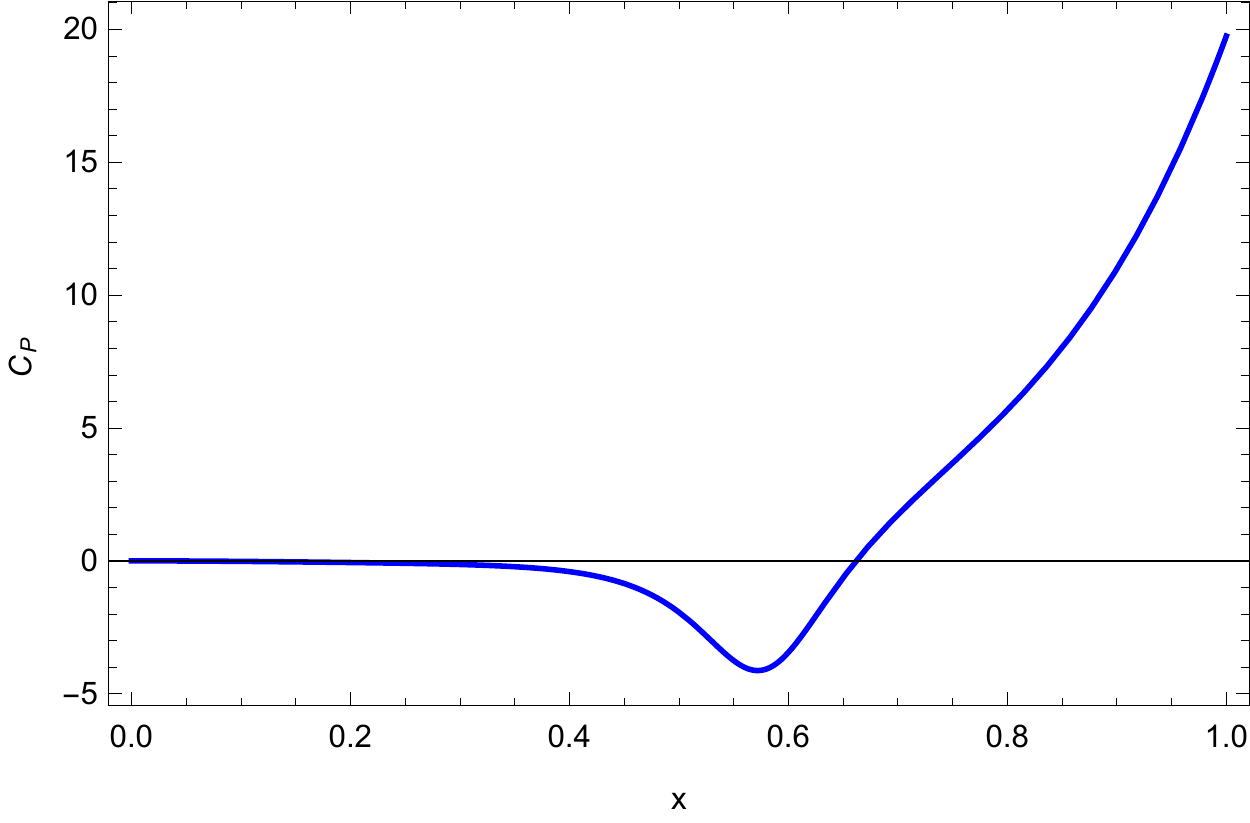,width=0.3\textwidth,angle=0,clip=true}}
 \subfigure[$Q^2=0.22$]{\epsfig{file=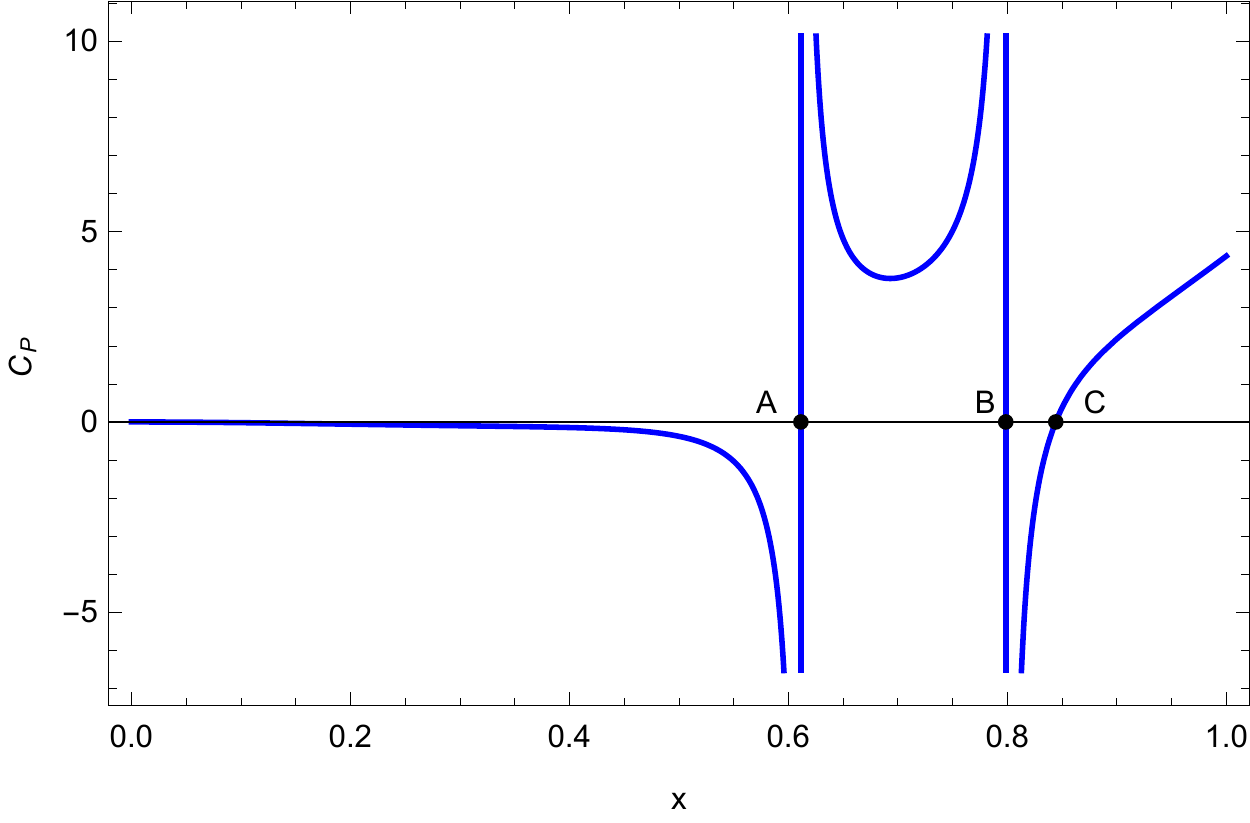,width=0.3\textwidth,angle=0,clip=true}}
 \subfigure[$Q^2=0.3$]{\epsfig{file=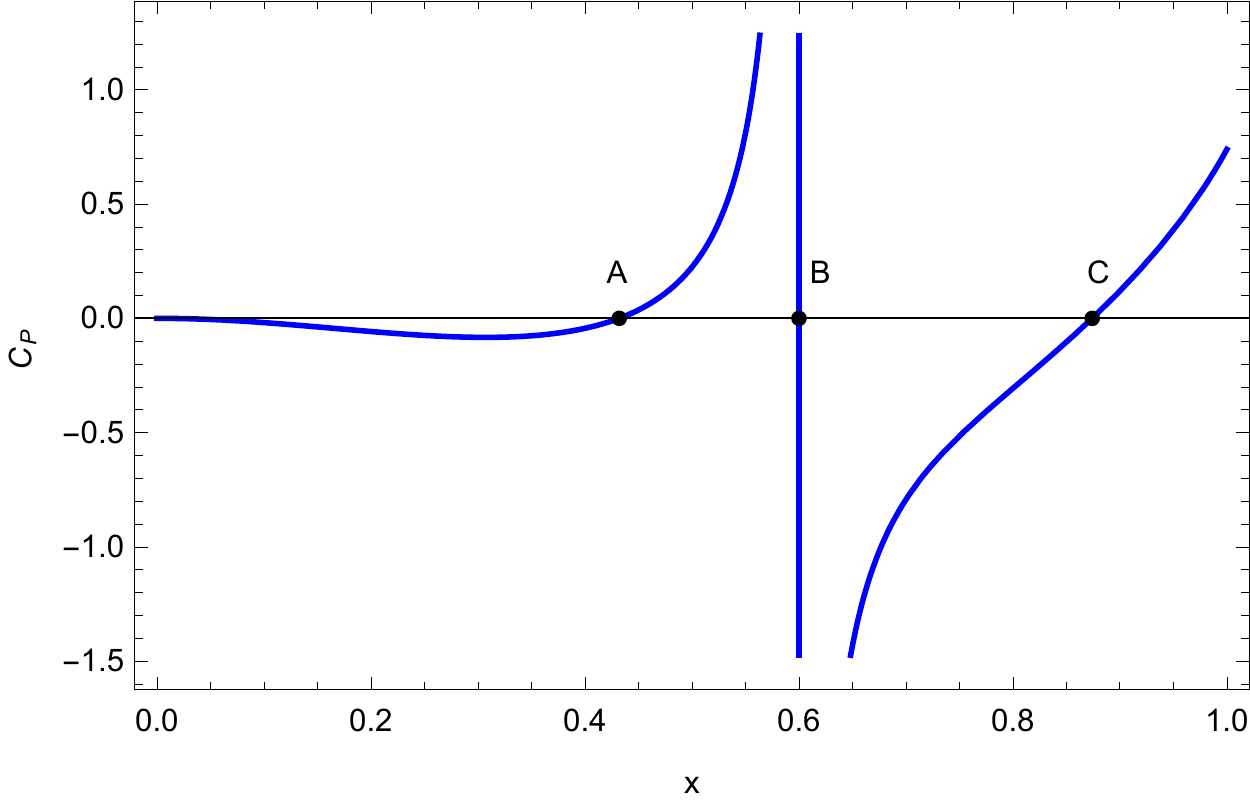,width=0.3\textwidth,angle=0,clip=true}}
\caption{The heat capacity $C_p$ as functions of $x$ for different $Q$.
\label{figcp}}
\end{center}
\end{figure}

To analyze the local thermodynamic stability of charged BHQ, we should also calculate the heat capacity at constant pressure,
\be\label{CPCBH}
C_p=\left.\frac{\partial{M}}{\partial{\tilde{T}_{eff}}}\right|_{Q,P_{eff}}=\tilde{T}_{eff}\left.\frac{\partial{S}}{\partial{\tilde{T}_{eff}}}\right|_{Q,P_{eff}}.
\ee
Its complete form is also given in the Appendix. For fixed $r_q=1$ , $C_p$ has five different behaviors according to different values of $Q$.  In detail, the five cases correspond to five intervals of $Q^2$, which are
$Q^2 \in (0,0.778]$,  $Q^2 \in (0.778,1/9]$, $Q^2 \in (1/9,0.195]$, $Q^2 \in (0.195,3/11]$ and $Q^2 \in (3/11,1/3]$. As is shown in Fig.\ref{figcp}, in cases (b) and (c), there is only one region where $C_p$ is positive. Thus, in these two cases the charged BHQ will be thermodynamically stable only in that region. In case (a) $C_p$ has three divergent points.  And in two disconnected regions "AB" and "CD", $C_p$ is positive. In case (d), there are two divergent points and two positive $C_p$ regions. In case (e), one divergent point and two positive $C_p$ regions. In fact, in case (e) the black hole is always thermodynamically stable because the requirement of positive temperature restricts the values of $x$ to be always on the right of point " C ", where $C_p$ is always positive. In cases (a) and (d), the two regions with positive $C_p$ is separated by a region with negative $C_p$. We want to know which one of the two regions with positive $C_p$ is more thermodynamically preferred. By this, we mean that if the black hole starts at the region labelled by "BC", will it transit to left region or right region? To judge the global thermodynamical stability, we need to consider Gibbs free energy.

\begin{figure}[ht]
\begin{center}
 \subfigure[$Q^2=0.06$]{\epsfig{file=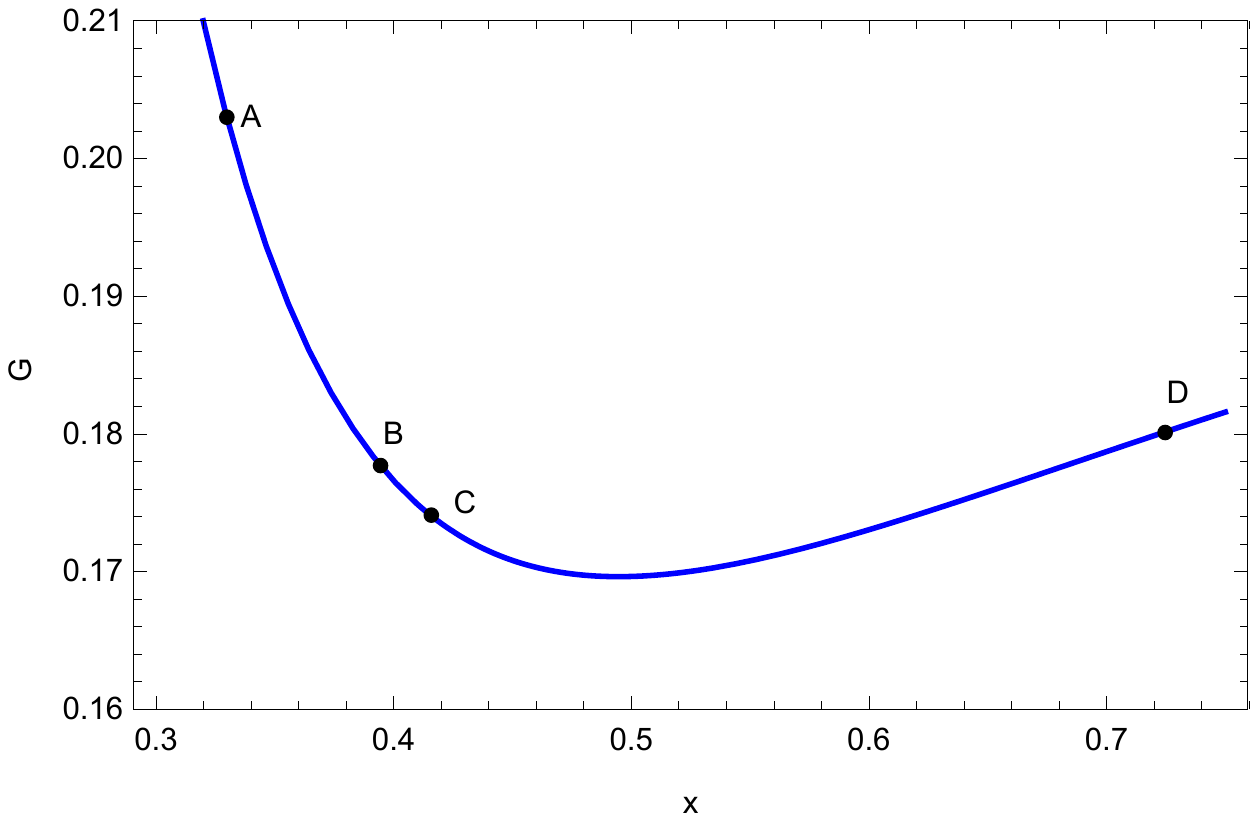,width=0.4\textwidth,angle=0,clip=true}}\hspace{0.5cm}
 \subfigure[$Q^2=0.22$]{\epsfig{file=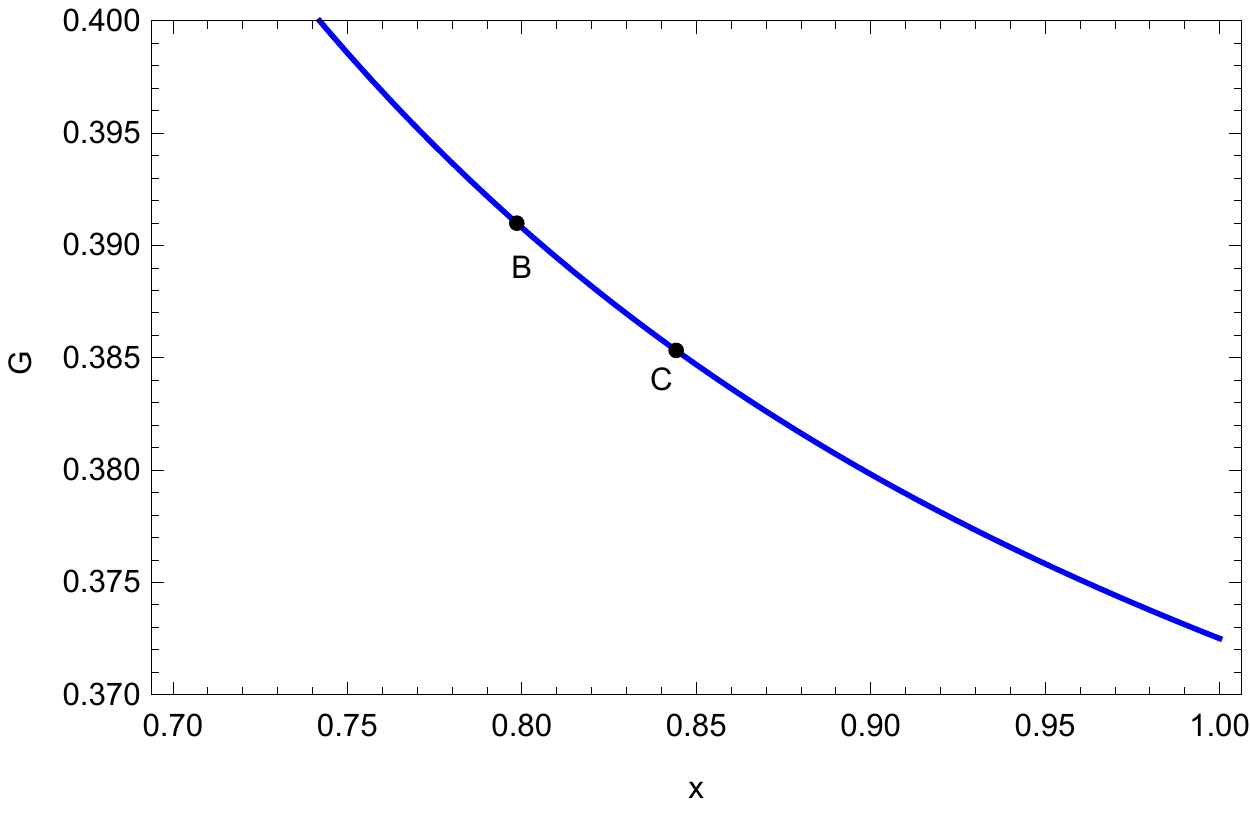,width=0.4\textwidth,angle=0,clip=true}}
\caption{The Gibbs free energy $G$ as functions of $x$ for $Q^2=0.06$ and $Q^2=0.22$.
\label{figGcp}}
\end{center}
\end{figure}

For the charged black hole surrounded by quintessence, Gibbs free energy can be defined as
\bea
G&=&M-T_{eff}S+P_{eff}V\no \\
&=&\frac{Q^2 \left(x^8+2 x^7+9 x^6+26 x^5+32 x^4+26 x^3+9 x^2+2 x+1\right)-x^8+3 x^6+8 x^5+3 x^4-x^2}{12 x^3 (x+1)^3},
\eea
where we have set $r_q=1$. As we all know, the thermodynamic state with lower Gibbs free energy is more stable. As is depicted in Fig.\ref{figGcp}, the region with positive $C_p$ on the right hand side of region "BC" is more thermodynamically preferred. From this we also know that, in cases (a) and (d), the charged BHQ will transit from a smaller black hole through an intermediate unstable black hole to a larger one. The phase transitions at the divergent points of $C_p$ are of second order.

\section{Conclusion}

In this paper, we have presented the global thermodynamic properties including thermodynamic stability and phase transition of uncharged and charged black holes surrounded by quintessence.
We choose the state parameter of quintessence as $\omega_q=-2/3$. With this, we find that the geometric structures of horizons of BHQ are similar to that of black holes in de Sitter space, although their metrics look different.
Not only that, due to the existence of multiple horizons, the BHQ also has the same problem as that in de Sitter black holes, namely, different temperatures for the different horizons generally.
Thus the  horizons cannot be in thermodynamic equilibrium, except for the Nariai or lukewarm case. We cannot study the thermodynamic system of BHQ globally, because the general formulation of non-equilibrium thermodynamics is still unknown at present. However, one can try to study this thermodynamic system globally in an effective way.

We constructed the effective thermodynamic system in two ways. First, we take the parameter $\alpha$ as a thermodynamic variable. From the effective first law of thermodynamics, we derived the effective thermodynamic quantities of uncharged and charged BHQ. It is shown that the uncharged BHQ is always thermodynamically unstable due to negative heat capacity. For the charged BHQ, we find that its effective temperature has the similar behavior to that of RNdS black hole\cite{Li-2016}. Specifically, for small electric charge $Q$, the effective temperature has a divergent point and is positive at two disconnected regions. However, critical behaviors of charged BHQ is different from that of RNdS black hole. In the small electric charge case, the heat capacity is negative for larger charged BHQ, while it is positive for larger RNdS black hole. Besides, there is only one divergent point for the charged BHQ and two divergent points for RNdS black hole. According to the heat capacity and free energy, we find that the charged BHQ will undergo a second-order phase transition at the point where heat capacity diverges. The another way is to take the volume between the black hole horizon and the quintessence horizon as a thermodynamic variable. This is done by analogy with de Sitter black holes. The thermodynamic quantities and thermodynamic properties are different in the two ways due to the different thermodynamic variables. The uncharged BHQ in the second way is also thermodynamically unstable because the heat capacity is always negative. For the charged case, it is interesting to find that the effective temperature behaves like that of RN black hole. By studying the heat capacity at constant volume $C_v$, we find that the charged BHQ can be thermodynamically stable only when the values of $(Q^2, x)$ are appropriately chosen, as is shown in Fig.\ref{figcvQ2x}. The behaviors of heat capacity at constant pressure $C_p$ are more complicated. $C_p$ has five different behaviors depending on the values of $Q$, shown in Fig.\ref{figcp}.
Combined with the Gibbs free energy, we find that the charged BHQ will undergo a second-order phase transition at the critical points and transit from a smaller black hole through an intermediate unstable black hole to a
larger one.

\section*{Acknowledgment}

This work is supported in part by the National Natural Science Foundation
of China (Grant No.11475108) and by the Doctoral Sustentation Fund of Shanxi Datong
University (2011-B-03).

\appendix

\section{Complete expressions of $C_v$ and $C_p$}

The complete forms of Eq.(\ref{CVCBH}) is:
\be
C_v=\frac{2 \pi  x^2 (x+1)^2 r_q^2 }{A(x)}\left[\left(x^6+x^2\right) r_q^2-Q^2 \left(x^6+2 x^5+2 x+1\right)\right],
\ee
where
\bea
A(x)&=&Q^2 \left(3 x^{10}+10 x^9+10 x^8-x^6-8 x^5-x^4+10 x^2+10 x+3\right)\no \\
&+&x^2 \left(x^8+4 x^7-6 x^4+4 x+1\right) r_q^2.
\eea
The complete forms of Eq.(\ref{CPCBH}) is:
\bea
C_p =-\frac{2 \pi  x^2 (x+1)^2 r_q^2}{B(x)} &\times&\left[Q^2 x^2 \left(-2 x^6+3 x^5+15 x^4+8 x^3+15 x^2+3 x-2\right) r_q^2 \right. \no \\
&+& \left.Q^4 \left(x^8+3 x^7-7 x^6-20 x^5-20 x^4-20 x^3-7 x^2+3 x+1\right) \right.\no \\
&+& \left.x^4 \left(x^4-4 x^3-4 x+1\right) r_q^4\right],
\eea
where
\bea
B(x)&=&-2 Q^2 x^2 \left(2 x^{10}+8 x^9+9 x^8-6 x^7-27 x^6-48 x^5-27 x^4-6 x^3+9 x^2+8 x+2\right) r_q^2 \no \\
&+& x^4 \left(x^8+4 x^7-4 x^5-14 x^4-4 x^3+4 x+1\right) r_q^4+Q^4 \left(3 x^{12}+16 x^{11}+30 x^{10} \right.  \no \\
&+& \left.20 x^9-30 x^8-132 x^7-210 x^6-132 x^5-30 x^4+20 x^3+30 x^2+16 x+3\right)
\eea

\bibliographystyle{JHEP}
\bibliography{H:/mms/References/references}

\end{document}